\documentclass[aps,prb,showpacs,twocolumn]{revtex4}

\usepackage{amsmath,amssymb,stmaryrd} 
\usepackage{color}
\usepackage{natbib}

\usepackage{ifpdf} 
\ifpdf 
  \usepackage[pdftex,breaklinks=true]{hyperref}
  \usepackage{graphicx} 
\else  
  \usepackage[hypertex,breaklinks=true]{hyperref}
  \usepackage[dvipdfmx]{graphicx} 
\fi

\renewcommand{\vec}[1]{\boldsymbol{#1}}
 
\renewcommand{\Im}{\mathop{\mathrm{Im}}\nolimits} 
\newcommand{\Log}{\mathop{\mathrm{Log}}\nolimits} 
\newcommand{\Arg}{\mathop{\mathrm{Arg}}\nolimits} 
\newcommand{\dilog}{\mathop{\mathrm{Li}_{2}}\nolimits} 
\newcommand{\sgn}{\mathop{\mathrm{sgn}}\nolimits}

\begin{document}

\title{Multi-terminal Anderson impurity model in nonequilibrium:
  Analytical perturbative treatment}
 
\author{Nobuhiko Taniguchi}

\affiliation{Institute of Physics, University of Tsukuba, Tennodai
  Tsukuba 305-8571, Japan}

\date{\today}

\begin{abstract}

We study the nonequilibrium spectral function of the single-impurity
Anderson model connecting with multi-terminal leads.  
The full dependence on frequency and bias voltage of the
nonequilibrium self-energy and spectral function is
obtained analytically up to the second-order perturbation regarding
the interaction strength $U$. 
High and low bias voltage properties are analyzed for a generic
multi-terminal dot, showing a crossover from the Kondo resonance to
the Coulomb peaks with increasing bias voltage. 
For a dot where the particle-hole symmetry is not present, we
construct a current-preserving evaluation of the nonequilibrium
spectral function for arbitrary bias voltage. 
It is shown that finite bias voltage does not split the Kondo
resonance in this order, and no specific structure due to
multiple leads emerges. Overall bias dependence is quite
similar to finite temperature effect for a dot with or without the
particle-hole symmetry.

\end{abstract}



\maketitle%

\section{Introduction}






Understanding strong correlation effect away from equilibrium has been
one of the most interesting yet challenging problems in condensed
matter physics. A prominent realization of such phenomena is embodied
in quantum transport through a nanostructure under finite bias
voltage.
To understand the interplay of the correlation effect and
nonequilibrium nature, nonequilibrium version of the single impurity
Anderson model (SIAM) and its extensions have been serving and
continue to do so as a central theoretical model.  The SIAM is indeed
considered to be one of the best studied strongly
correlated model, and despite its apparent simplicity, it exhibits rich
physics already in equilibrium, such as the Coulomb blockade and the
Kondo physics that have been observed in experiments.
Equilibrium properties of the SIAM have been well understood thanks to
concerted efforts of several theoretical approaches over years --- by
perturbative treatment, Fermi liquid description, as well as exact
results by the Bethe ansatz method and numerical renormalization group
(NRG) calculations~(see for instance,
\onlinecite{Tsvelick83b,HewsonBook93,Bulla08}.)
In contrast, the situation of the nonequilibrium SIAM is not so
satisfactory. Each of the above approaches has met some difficulty in
treating nonequilibrium phenomena.  A theoretical approach that can
deal with the strong correlation effect in nonequilibrium is still
called for.

Notwithstanding, a number of analytical and numerical methods have been
devised to investigate nonequilibrium stationary phenomena:
nonequilibrium perturbation
approaches~\cite{Hershfield92,Fujii03,Fujii05,Hamasaki07,Muhlbacher11} and its
modifications~\cite{Yeyati93,Yeyati99,Aligia06}, the non-crossing
approximation~\cite{Meir93}, the functional renormalization
group treatment~\cite{Schmidt10},
quantum Monte Carlo calculations on the Keldysh
contour~\cite{Schmidt08,Werner10}, the iterative real-time path
integral method~\cite{Weiss08} and so on.  Unfortunately those approaches
fail to give a consistent picture concerning the finite bias effect on
the dot spectral function, particularly regarding a possible
splitting of the Kondo resonance.

As for the equilibrium SIAM, the second-order perturbation regarding
the Coulomb interaction $U$ on the
dot~\cite{Yamada75,Yamada75b,Yoshida75,Zlatic83} is known to capture
essential features of Kondo physics and agrees qualitatively well with
exact results obtained by the Bethe ansatz and NRG
approaches~\cite{Tsvelick83b,HewsonBook93,Zlatic83}. Such good
agreement seems to persist in nonequilibrium stationary state at
finite bias voltage.
For the two-terminal particle-hole (PH) symmetric SIAM, a recent study
by \citet{Muhlbacher11} showed that the nonequilibrium second order
perturbation calculation of the spectral function agrees with that
calculated by the diagrammatic quantum Monte Carlo simulation,
excellently up to interaction strength $U \sim 2\gamma$ (where
$\gamma$ is the total relaxation rate due to leads); pretty well even
for $U \lesssim 8\gamma$ at bias voltage $eV \lesssim 2\gamma$. A
typical magnitude of $U/\gamma$ of a semiconductor quantum dot is
roughly $1\sim 10$ depending on the size and the configuration of the
dot. Therefore, there is a good chance of describing a realistic
system within the validity of nonequilibrium perturbation approach.


The great advantage of semiconductor dot systems is to allow us to
control several physical parameters. Those include changing gate
voltage as well as configuring a more involved structure such as 
a multi-terminal dot~\cite{Sun01,Lebanon01,Cho03,Sanchez05,Shah06,Leturcq05} or an interferometer
embedding a quantum dot. 
Theoretical treatments often limit themselves to a system with the PH
symmetry where the dot occupation number is fixed to be one half per
spin.  Although assuming the PH symmetry makes sense and comes in
handy in extracting the essence of the Kondo resonance, we should bear
in mind that such symmetry is not intrinsic and can be broken easily
in realistic systems, by gate voltage, asymmetry of the coupling with
the leads, or asymmetric drops of bias
voltage~\cite{Aligia12,Munoz13}.  The PH asymmetry commonly appears in
a multi-terminal dot or in an interferometer embedding a quantum dot.
It is also argued that the effect of the PH asymmetry might be
responsible for the deviation observed in nonequilibrium
transport experiments from the ``universal'' behavior of the PH
symmetric SIAM~\cite{Aligia12,Munoz13}.  To work on realistic
systems, it is imperative to understand how the PH asymmetry affects
nonequilibrium transport.


In this paper, we examine the second-order nonequilibrium perturbation
regarding the Coulomb interaction $U$ of the multi-terminal SIAM. The
PH symmetry is not assumed, and miscellaneous types of asymmetry of
couplings to the leads and/or voltage drops are incorporated as a
generic multi-terminal configuration. Our main focus is to provide
solid analytical results of the behavior of the nonequilibrium
self-energy and hence the dot spectral function for the full range of
frequency and bias voltage, within the validity of the second-order
perturbation theory of $U$.  The result encompasses Fermi liquid
behavior as well as incoherent non-Fermi liquid contribution, showing
analytically that increasing finite bias voltage leads to a crossover
from the Kondo resonance to the Coulomb blockade behaviors.
The present work contrasts preceding perturbative
studies~\cite{Hershfield92,Hamasaki04,Hamasaki07,Fujii03,Fujii05}
whose evaluations relied on either numerical means or a
small-parameter expansion of bias voltage and frequency.  The only
notable exception, to the author's knowledge, is a recent work by
\citet{Muhlbacher11}, which succeeded in evaluating analytically the
second-order self-energy for the two-terminal PH symmetric dot.
Intending to apply such analysis to a wider range of realistic systems
and examine the effect that the two-terminal PH symmetric SIAM cannot
capture, we extend their approach to a generic multi-terminal dot
where the PH symmetry may not necessarily be present.

An embarrassing drawback of using the nonequilibrium perturbation
theory is that when one has it naively apply to the PH asymmetric SIAM,
it may disrespect the preservation of the steady
current.~\cite{Hershfield92}  As a result, one then needs some 
current-preserving prescription, and different self-consistent schemes
have been proposed and 
adopted.~\cite{Yeyati93,Matsumoto00,Aligia06} 
As will be seen, the current preserving condition involves all the
frequency range, not only of low frequency region that validates Fermi
liquid description but also of the incoherent non-Fermi liquid part
(See Eq.~\eqref{eq:self-consistency} below).  Therefore, an
approximation based on the low energy physics, particularly the Fermi
liquid picture, should be used with care.
The self-energy we will construct analytically is checked to
satisfy the spectral sum rule at finite bias voltage, so that we
regard it as giving a consistent description for the full range of
frequency in nonequilibrium. By taking its advantage, we also demonstrate a
self-consistent, current-preserving calculation of the nonequilibrium
spectral function for a system where the PH symmetry is not present.

The paper is organized as follows. In Sec.~\ref{sec:model}, we
introduce the multi-terminal SIAM in nonequilibrium.  We review
briefly how to obtain the exact current formula by clarifying the role
of the current conservation at finite bias voltage.
Section~\ref{sec:self-energy} presents analytical expression of the
retarded self-energy for a general multi-terminal dot up to the
second-order of the interaction strength.  Subsequently in
Sec.~\ref{sec:discussion}, we examine and discuss its various analytical
behaviors including high- and low-bias voltage limits.
Sec.~\ref{sec:spectral-fn} is devoted to constructing a
non-equilibrium spectral function using the self-energy obtained in
the previous section. We focus our attention on two particular situations:
(1)~self-consistent, current-preserving evaluation of the
nonequilibrium spectral function for the two-terminal PH asymmetric
SIAM, and (2)~multi-terminal effect of the PH symmetric SIAM.  Finally
we conclude in Sec.~\ref{sec:conclusion}.
Mathematical details leading to our main analytical result
Eq.~\eqref{eq:SigmaR} as well as other necessary material regarding
dilogarithm is summarized in Appendices.

\section{The multi-terminal Anderson impurity model and the current formula}
\label{sec:model}

\subsection{Model}

The model we consider is the single impurity Anderson model connecting
with multiple leads $a=1, \ldots, N$ whose chemical potentials
are sustained by $\mu_{a}$.  The total Hamiltonian of the system
consists of $H = H_{D} + H_{T} + \sum_{a} H_{a}$, where $H_{D}$,
$H_{T}$ and $H_{a}$ represent the dot Hamiltonian with the Coulomb
interaction, the hopping term between the dot and the leads, and the
Hamiltonian of a noninteracting lead $a$, respectively.
They are specified by
\begin{align}
& H_{D} = \sum_{\sigma} \epsilon_{d}\, n_{\sigma} + U n_{\uparrow}
n_{\downarrow},\\ 
& H_{T} = \sum_{a,\sigma} \left( V_{d a}\, d^{\dagger}_{\sigma}
  c_{a\vec{k}\sigma} + V_{ad}\, c^{\dagger}_{a\vec{k}\sigma} d_{\sigma} \right),
\end{align}
where $n_{\sigma} = d^{\dagger}_{\sigma} d_{\sigma}$ is the dot
electron number operator with spin $\sigma$ and $c_{a\vec{k}\sigma}$
are electron operators at the lead $a$.  In the following, we consider
the spin-independent transport case, but an extension to the
spin-dependence situation such as in the presence of magnetic field or
ferromagnetic leads is straightforward.
When applying the wide-band limit, all the effect of the lead $a$ is
encoded in terms of its chemical potential $\mu_{a}$ and 
relaxation rate $\gamma_{a} = \pi |V_{da}|^{2}
\rho_{a}$, where $\rho_{a}$ is the density of
states of the lead $a$.  
The dot level $\epsilon_{d}$ controls the average occupation number on
the dot; it corresponds roughly to $2$, $1$, $0$ for $\epsilon_{d}
\lesssim -U$, $-U \lesssim \epsilon_{d} \lesssim 0$, and $0 \lesssim
\epsilon_{d}$, respectively. The PH symmetry is realized when
$\epsilon_{d} =-U/2$ and $\langle n_{\sigma} \rangle = 1/2$ (see
Eqs.~\eqref{eq:self-consistency} and \eqref{eq:HF-self-consistency}
below).

\subsection{Multi-terminal current and current conservation} 
\label{sec:current-formula}

We here briefly summarize how the current through the dot is
determined in a multi-terminal setting.  Special attention is paid
to the role of the current conservation because it has been known that
nonequilibrium perturbation calculation does not respect it in
general.~\cite{Hershfield92}  We
illustrate how to ensure the current conservation by requiring
minimally.  The argument below is valid regardless of a specific
approximation scheme chosen, whether nonequilibrium perturbation or
any other approach.

Following the standard protocol of the Keldysh
formulation~\cite{Meir92}, we start with writing the current $I_{a}$
flowing from the lead $a$ to the dot in terms of the dot's lesser
Green function $G^{-+}_{\sigma}$ and the retarded one
$G^{R}_{\sigma}$:
\begin{align}
& I_{a} 
= -\frac{e}{\pi\hbar}\sum_{\sigma} \int d \omega \left[ i
  \gamma_{a}\, G^{-+}_{\sigma}(\omega) - 
2 \gamma_{a} f_{a} \Im G^{R}_{\sigma}(\omega)\, \right],
\end{align}
where $f_{a}(\omega) = 1/(e^{\beta(\omega-\mu_{a})}+1)$ is the Fermi
distribution function at the lead $a$.
As the present model preserves the total spin as well as charge, the
net spin current flowing to the dot should 
vanish in the steady state, which imposes the integral relation
between $G^{-+}_{\sigma}$ and $G^{R}_{\sigma}$,
\begin{align}
  & \int^{\infty}_{-\infty} d\omega \left[ i \gamma\, G^{-+}_{\sigma}
    (\omega) +2 \gamma \bar{f}(\omega) \Im
    G^{R}_{\sigma} (\omega) \right] = 0.
\label{eq:Gless-GR}
\end{align}
Here we have introduced the total relaxation rates $\gamma = \sum_{a}
\gamma_{a}$ and the effective Fermi distribution $\bar{f}$ weighted by
the leads,
\begin{align}
  \bar{f}(\omega) = 
  \sum_{a} \frac{\gamma_{a}}{\gamma}\, f_{a}(\omega).
\label{eq:effective-f}
\end{align}
When we ignore the energy dependence of the relaxation rates $\gamma_{a}$,
we can recast Eq.~\eqref{eq:Gless-GR} into a more familiar form
\begin{align}
&  n_{\sigma} = -\frac{1}{\pi} \int^{\infty}_{-\infty} d \omega\, 
  \bar{f}(\omega) \Im G^{R}_{\sigma} (\omega),
\label{eq:self-consistency}
\end{align}
because $2i \pi n_{\sigma} = \int d\omega G^{-+}_{\sigma}(\omega)$ is
the definition of the exact dot occupation number. Note the quantity
$-\Im G^{R}_{\sigma}(\omega)/\pi$ is nothing but the exact
dot spectral function out of equilibrium.  We emphasize that
Eq.~\eqref{eq:Gless-GR} (or equivalently
Eq.~\eqref{eq:self-consistency}) is the minimum, exact requirement that
ensures the current preservation.  It constrains the exact $G^{-+}$
and $G^{R}$ that depend on the interaction as well as bias voltage in
a nontrivial way.  One can accordingly eliminate $\int d\omega\,
G^{-+}(\omega)$ in $I_{a}$, to reach the Landauer-Buttiker type current
formula at the lead $a$,
\begin{align}
  I_{a} = -\frac{e}{\pi \hbar} \sum_{b,\sigma}
  \frac{\gamma_{a} \gamma_{b}}{\gamma}
  \int d\omega\, (f_{b} - f_{a}) \Im G^{R}_{\sigma}
(\omega).
\end{align}
Or, the current conservation allows us to write it as
\begin{align}
& I_{a} 
= \frac{e\gamma_{a}}{\hbar} 
\sum_{\sigma} \left[n_{\sigma} - \mathcal{N}_{\sigma} (\mu_{a}) \right],
\end{align}
where $\mathcal{N}_{\sigma}(\varepsilon)$ is the exact number of
states with spin $\sigma$ at finite temperature in general, defined by
\begin{align}
& \mathcal{N}_{\sigma}(\mu) 
= -\frac{1}{\pi} \int d\varepsilon\,
\frac{\Im G^{R}_{\sigma}(\varepsilon)}{e^{\beta(\varepsilon- \mu)} + 1}.
\end{align}
It tells us that differential conductance $\partial I_{a}/\partial
\mu_{a}$ with fixing all other $\mu$'s is proportional to the
nonequilibrium dot spectral function, provided changing
$\mu_{a}$ does not affect the occupation
number~\cite{Sun01,Lebanon01,Cho03,Sanchez05}. Such situation is
realized, for instance,  when a probe lead couples weakly to the dot.

The case of a noninteracting dot always satisfies the current
preserving condition Eq.~\eqref{eq:Gless-GR} as $G^{-+}_{\sigma}
(\omega) = -2i\bar{f}(\omega) \Im G^{R}_{\sigma}(\omega)$ holds for
any $\omega$; the distribution function of dot electrons
$f_{\text{dot}}(\omega) = G^{-+}(\omega)/(2i\pi)$ is equal to
$-\bar{f}(\omega) \Im G^{R}_{\sigma}(\omega)/\pi$.
This is not the case for an interacting dot, however.
As for the interacting case, not so much can be said.  We only see the
special case with the two-terminal PH symmetric dot satisfy
Eq.~\eqref{eq:self-consistency} by choosing $n_{\sigma}=1/2$
irrespective of interaction strength. Except for this PH symmetric
case, a general connection between $G^{-+}$ and $G^{R}$ is not known
so far.  It is remarked that, based on the quasiparticle picture, a
noninteracting relation $G^{-+}_{\sigma} (\omega) = -2i\bar{f}(\omega)
\Im G^{R}_{\sigma}(\omega)$ is sometimes used to deduce an approximate
form of $G^{-+}$ out of $G^{R}$ for an interacting dot.  Such
approximation is called the Ng's ansatz~\cite{Ng93,Dinu07}.  Although
it might be be simple and handy, its validity is far
from clear. We will not rely on such additional approximation below.
It is also important to distinguish in
Eq.~\eqref{eq:self-consistency} the electron occupation number
$n_{\sigma}$ from the quasiparticle occupation number
$\tilde{n}_{\sigma}$, as the two quantities are different at finite
bias voltage since the Luttinger relation holds only in
equilibrium~\cite{Luttinger60}. 
Contribution to the dot occupation number comes from all range of
frequency, including the incoherent part.
One sees fulfilling the spectral weight sum rule
$-\int^{\infty}_{-\infty} d\omega \Im
G^{R}(\omega) /\pi= 1$ crucial to have the dot occupation number
$n_{\sigma}$ normalized correctly.  
In general, one needs to determine $n_{\sigma}$ appropriately to
satisfy Eq.~\eqref{eq:self-consistency} as a function of interaction
and chemical potentials of the leads. The applicability of
quasi-particle approaches that ignores the incoherent part is unclear.

\section{Analytical evaluation of the self-energy}
\label{sec:self-energy}

In this section, we evaluate analytically the nonequilibrium retarded
self-energy up to the second order of interaction strength $U$ for the
multi-terminal SIAM.  We first examine the contribution at the
first-order and the role of current preservation. Then we present the
analytical result of the second-order self-energy in terms of
dilogarithm.

Following the standard treatment of the Keldysh
formulation,~\cite{LifshitzBook81} the nonequilibrium Green function
and the self-energy take a matrix structure
\begin{align}
& \hat{G}  =
\begin{pmatrix}
  G^{--} & G^{-+} \\
  G^{+-} & G^{++} 
\end{pmatrix};
\quad
\hat{\Sigma}  =
\begin{pmatrix}
  \Sigma^{--} & \Sigma^{-+} \\
  \Sigma^{+-} & \Sigma^{++} 
\end{pmatrix},
\label{eq:G-and-Sigma}
\end{align}
satisfying symmetry relations $G^{--} + G^{++} = G^{-+} + G^{+-}$ and
$\Sigma^{--} + \Sigma^{++} = - \Sigma^{-+} - \Sigma^{+-}$.  The retarded Green function is defined by $G^{R} = G^{--} +
G^{-+}$; the retarded self-energy, by $\Sigma^{R} = \Sigma^{--} +
\Sigma^{-+}$.

\begin{figure}
  \centering
  \includegraphics[width=0.3\linewidth,bb=0 0 83 106,clip]{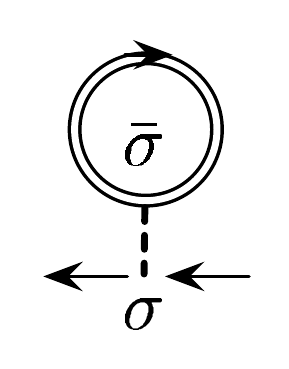}
  \caption{The Hartree-type contribution of the self-energy $U\tau_{3}\,
    n_{\bar{\sigma}} = \pm U n_{\bar{\sigma}}$. The 
    double line refers to the exact Green function.}
  \label{fig:Hartree-type}
\end{figure}

To proceed evaluation, it is convenient to classify self-energy
diagrams into the two types: the Hartree-type diagram
(Fig.~\ref{fig:Hartree-type}) that can be disconnected by cutting a
single interaction line, and the rest which we call the correlation
part and reassign the symbol $\Sigma$ to.  The latter starts at the
second-order. The resulting Green function (matrix) takes a form of
\begin{align}
\hat{G}_{\sigma}(\omega) = \left[ \hat{G}_{0\sigma}^{-1}(\omega) - U
  \tau_{3}\, n_{\bar{\sigma}}- \hat{\Sigma}_{\sigma}(\omega) \right]^{-1},
\end{align}
where $\tau_{3}$ represents a Pauli matrix of the Keldysh structure,
and $n_{\bar{\sigma}}$ refers to the exact occupation
number of the dot electron with the opposite spin. 
Accordingly, the corresponding retarded Green function becomes
\begin{align}
G^{R}_{\sigma}(\omega) = \frac{1}{\omega -
    E_{d\sigma} + i\gamma - \Sigma_{\sigma}^{R}(\omega)},
\label{eq:GR}
\end{align}
where $E_{d\sigma} = \epsilon_{d} + U n_{\bar{\sigma}}$ is the Hartree
level of the dot.  

\subsection{Current preservation at the first order}

Before starting evaluating the correlation part $\Sigma^{R}$ that
starts contributing at the second order, it is worthwhile to 
examine the current preserving condition
Eq.~\eqref{eq:self-consistency} at the first order.  
At this order, it reduces to the self-consistent Hartree-Fock equation 
for the dot occupation number $n_{\sigma}^{0}$:
\begin{align}
& n_{\sigma}^{0}
= \frac{1}{2} + \frac{1}{\pi}  \sum_{a} \frac{\gamma_{a}}{\gamma}
\arctan \left[\frac{\mu_{a} - \epsilon_{d} - U n_{\bar{\sigma}}^{0}}{\gamma}
  \right].
\label{eq:HF-self-consistency}
\end{align}
It shows how the two-terminal PH symmetric SIAM is special by
choosing $\epsilon_{d}+U/2=0$, $\gamma_{a} = \gamma/2$ and $\mu_{a} = \pm eV/2$; the second term of the right-hand side
vanishes by having the solution $n^{0}_{\sigma}=1/2$ even at finite bias
voltage.
It also indicates that the current preservation necessarily has the
occupation number depend on asymmetry of the lead couplings as well as
interaction strength for the PH asymmetric SIAM.  Indeed, for a
small deviation from the PH symmetry and bias voltage, we see
the Hartree-Fock occupation number behave as
\begin{align}
  n_{\sigma}^{0} - \frac{1}{2}
\approx \frac{\bar{\mu} - \epsilon_{d} -
  U/2}{\pi\gamma} \left( 1- \frac{U}{\pi \gamma} + \cdots \right),
\end{align}
where $\bar{\mu}$ is the average chemical potential weighted by leads,
\begin{align}
  \bar{\mu} = \sum_{a} \frac{\gamma_{a}}{\gamma} \mu_{a}.
\label{def:mubar}
\end{align}
Note $\bar{\mu}$ vanishes when no bias
voltage applies, as we incorporate the overall net offset by leads
into $\epsilon_{d}$. 

\subsection{Analytical evaluation of the correlation part of the
  self-energy}

\begin{figure}
  \centering
  \includegraphics[width=0.55\linewidth,bb=0 50 200 190,clip]{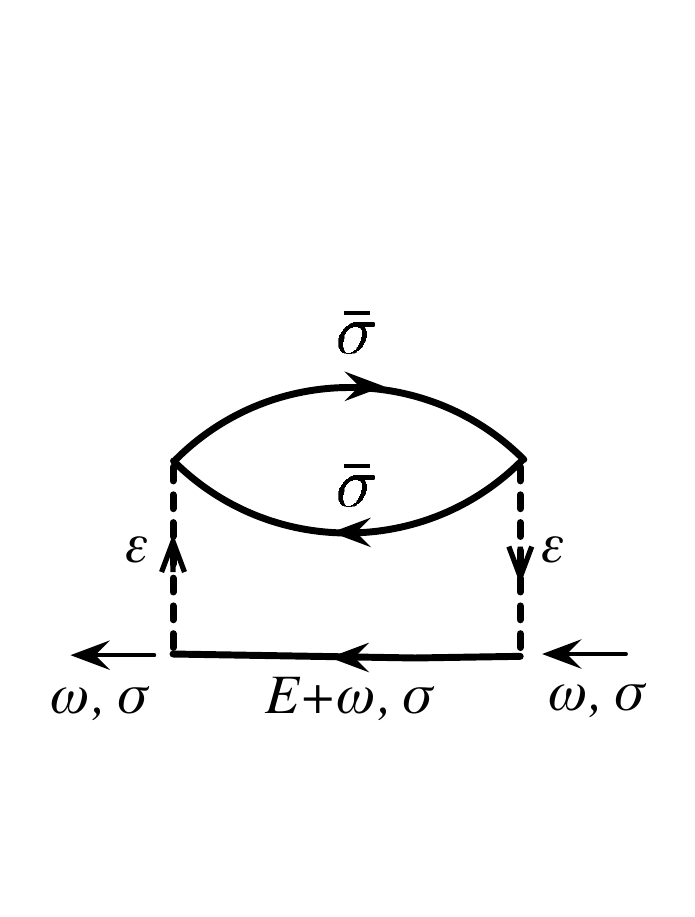}
  \caption{The correlation part of the self-energy at the second-order
    contribution. }
  \label{fig:Sigma2}
\end{figure}

\begin{figure}
  \centering
  \includegraphics[width=0.65\linewidth,bb=0 80 209 180,clip]{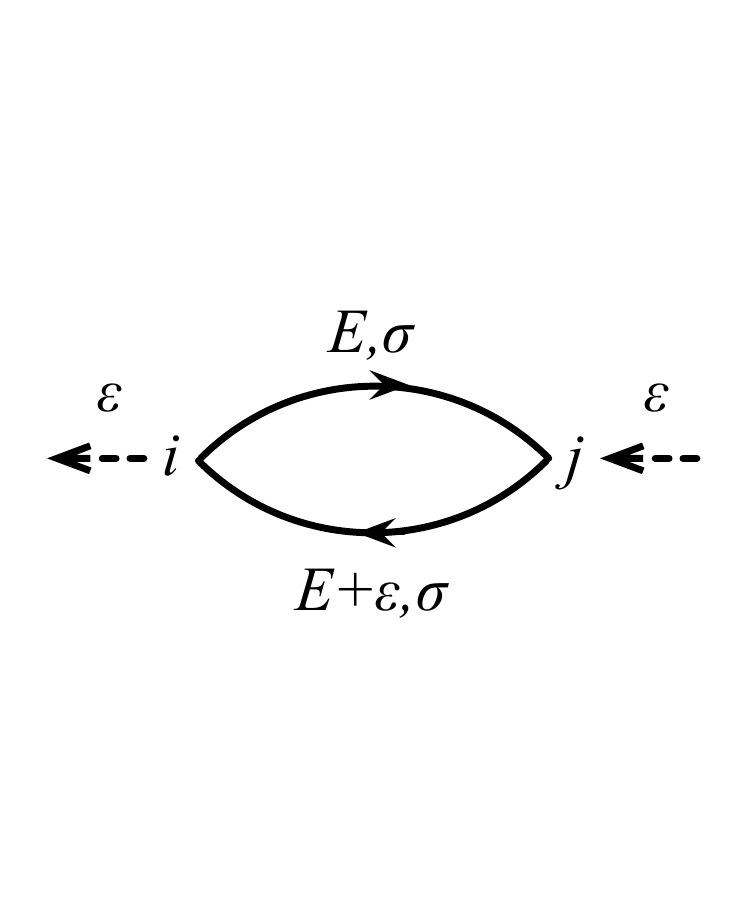}
  \caption{The polarization part}
  \label{fig:Pi}
\end{figure}

Following the standard perturbation treatment of the Keldysh
formulation, we see there is only one diagram contributing to
$\Sigma^{R}_{\sigma}$ at the second order (Fig.~\ref{fig:Sigma2})
after eliminating the Hartree-type contribution. The contribution is
written as
\begin{align}
& \hat{\Sigma}(t_{1},t_{2}) 
= - i \hbar U^{2} 
\begin{pmatrix}
  G^{--}_{12} \Pi^{--}_{21} & - G^{-+}_{12} \Pi^{+-}_{21} \\
  - G^{+-}_{12} \Pi^{-+}_{21} & G^{++}_{12} \Pi^{++}_{21}
\end{pmatrix}
\end{align}
where $G^{ij}_{12}= G^{ij}(t_{1},t_{2})$ refer to to the unperturbed
Green functions (including the Hartree term), whose concrete
expressions are found in Appendix A.  The polarization matrix
$\hat{\Pi}$ is defined by $\Pi^{ij}_{12} = i\hbar\, G^{ij}_{12}
G^{ji}_{21}$ (Fig.~\ref{fig:Pi}).\footnote{We define the polarization to satisfy the
  symmetric relation $\Pi^{--} + \Pi^{--}=\Pi^{-+}+\Pi^{+-}$.}

As was shown by the current formula in the previous section, we need
only the dot spectral function to study quantum transport, hence
$\Sigma^{R}$ suffices. Therefore it is more advantageous to work
on the RAK-representation, where the polarization parts become
\begin{subequations}
  \begin{align}
    & \Pi^{R}_{12} = \frac{i\hbar}{2} \left[ G^{R}_{12} G^{K}_{21} +
      G^{K}_{12} G^{A}_{21}
    \right], \\
    & \Pi^{A}_{12} = \frac{i\hbar}{2} \left[ G^{A}_{12} G^{K}_{21} +
      G^{K}_{12} G^{R}_{21}
    \right], \\
    & \Pi^{K}_{12} = \frac{i\hbar}{2} \left[ G^{K}_{12} G^{K}_{21} +
      G^{R}_{12} G^{A}_{21} + G^{A}_{12} G^{R}_{21} \right],
  \end{align}
\label{eq:def-Pi}
\end{subequations}
and their Fourier transformations are given in Appendix B.
Accordingly, we can express the
retarded self-energy $\Sigma^{R}$ as
\begin{align}
\Sigma^{R}_{\sigma} (\omega) =  -\frac{iU^{2}}{4\pi} \Big[
I_{1} (\omega) + I_{2}(\omega) \Big],
\end{align}
where
\begin{align}
    & I_{1}(\omega) = \int^{+\infty}_{-\infty} dE\,
    G^{R}_{\sigma}(E+ \omega) \Pi^{K}_{\bar{\sigma}}(E), 
    \label{eq:I1}\\
    & I_{2}(\omega) = \int^{+\infty}_{-\infty} dE\, G^{K}_{\sigma}(E+
    \omega) \Pi^{A}_{\bar{\sigma}}(E). \label{eq:I2}
\end{align}
The above second-order expression of $\Sigma^{R}$ is standard, but it
has so far been mainly used for numerical evaluation, quite often
restricted for the two-terminal PH symmetric SIAM. We intend to
evaluate Eqs.~(\ref{eq:I1},\ref{eq:I2}) analytically for the generic
multi-terminal SIAM, along the line employed in
Ref.~\onlinecite{Muhlbacher11}.

Delegating all the mathematical details to Appendices C and D, we 
summarize our result of the analytical evaluation of $\Sigma^{R}$ as follows:
\begin{widetext}
\begin{align}
& \Sigma^{R}_{\sigma}(\omega) = \frac{i\gamma U^{2}}{8\pi^{2}
  (\omega - E_{d\sigma} + i\gamma)}
\bigg[ \frac{\Xi_{1}(\omega - E_{d\sigma})}{\omega - E_{d\sigma} - i\gamma} 
+ \frac{\Xi_{2}(\omega - E_{d\sigma})}{\omega - E_{d\sigma} + 3i\gamma} 
+ \frac{\Xi_{3}}{2i\gamma} \bigg],
\label{eq:SigmaR}
\end{align}
Here, functions $\Xi_{i}$ ($i=1,2,3$) are found to be
(using $\zeta_{a \sigma} = \mu_{a} - E_{d\sigma}$),
\begin{subequations}
\label{eq:Xi123}
\begin{align}
\Xi_{1}(\varepsilon) &= \tfrac{2\pi^{2} \varepsilon}{i\gamma} +
      \sum_{a,b,\beta} \tfrac{4\gamma_{a}
        \gamma_{b}}{\gamma^{2}} \Big[ \dilog(\tfrac{-
        \varepsilon + \zeta_{a\sigma}}{\beta
        \zeta_{b\bar{\sigma}} + i\gamma}) +
      \dilog(\tfrac{-\varepsilon - \beta \zeta_{b\bar{\sigma}}
      }{-\zeta_{a\sigma} +i\gamma}) + \tfrac{1}{2} \Log^{2}
      (\tfrac{-\zeta_{a\sigma} + i\gamma}{\beta \zeta_{b\bar{\sigma}} +
        i\gamma}) \Big] %
\nonumber \\ & \qquad \qquad 
+ \sum_{a,b,\beta}
      \tfrac{4\gamma_{a}
        \gamma_{b}}{\gamma^{2}} \Big[ \dilog
      (\tfrac{-\varepsilon+ \beta \zeta_{a\bar{\sigma}} }{\beta
        \zeta_{b \bar{\sigma}} + i\gamma}) + \tfrac{1}{4} \Log^{2}
      (\tfrac{-\zeta_{a\bar{\sigma}}+i\gamma}{\zeta_{b\bar{\sigma}} +
        i\gamma})
      \Big], \\ %
 \Xi_{2}(\varepsilon) &=  6\pi^{2}
      - \sum_{a,b,\beta} \tfrac{4\gamma_{a} \gamma_{b}}{\gamma^{2}} \Big[
      \Lambda(\tfrac{\varepsilon -\zeta_{a\sigma} +
        2i\gamma}{\beta \zeta_{b\bar{\sigma}} + i\gamma}) +
      \Lambda(\tfrac{\varepsilon - \beta \zeta_{b\bar{\sigma}}
        + 2i\gamma}{\zeta_{a\sigma} +i\gamma}) + \tfrac{1}{2} \Log^{2}
      (\tfrac{\zeta_{a\sigma} + i\gamma}{\beta \zeta_{b \bar{\sigma}} +
        i\gamma}) \Big] %
\nonumber \\ & \qquad  \qquad
- \sum_{a,b,\beta}
      \tfrac{4\gamma_{a} \gamma_{b}}{\gamma^{2}} \Big[ \Lambda
      (\tfrac{\varepsilon+ \beta \zeta_{a\bar{\sigma}} +
        2i\gamma }{\beta \zeta_{b\bar{\sigma}} + i\gamma}) +
      \tfrac{1}{4} \Log^{2} ( \tfrac{-\zeta_{a\bar{\sigma}} +
        i\gamma}{\zeta_{b\bar{\sigma}} + i\gamma} )
      \Big], \\ %
 \Xi_{3} &= \left[\sum_{a}
      \tfrac{2\gamma_{a}}{\gamma}
      \Log \left( \tfrac{-\zeta_{a\bar{\sigma}} +
          i\gamma}{\zeta_{a\bar{\sigma}} + i\gamma}\right) \right]^{2},
    \end{align}
\label{eq:Xi-123}
  \end{subequations}
\end{widetext}
where the summations over $\beta = \pm 1$ as well as terminals $a,b$
are understood. Function $\dilog(z)$ is dilogarithm, whose definition
as well as various useful properties are summarized in Appendix C;
$\Lambda(z)$ is defined by~\footnote{The definition of $\Lambda(z)$ is
  equivalent to that given in Ref.~\onlinecite{Muhlbacher11}, but we
  prefer writing it in this form because its analyticity is more
  transparent.}
\begin{align}
  \Lambda(z) =  \dilog(z)  + \left[ \Log(1-z) - \Log(z-1) \right] \Log
  z.
\label{def:Lambda}
\end{align}
Analytical formula $\Sigma^{R}$ given in Eqs.~\eqref{eq:SigmaR} and
\eqref{eq:Xi-123} constitutes the main result of the present paper.
Consequently, the nonequilibrium spectral function of the multi-terminal
SIAM is given analytically for full range of frequency and bias
voltage, once one chooses $n_{\sigma}$ to satisfy
Eq.~\eqref{eq:self-consistency}. The result also applies to a more
involved structured system, such as an
interferometer embedding a quantum dot, by simply replacing
$\epsilon_{d}$ and $\gamma_{a}$ to take account of those geometric effect.

\section{Various analytical behaviors}
\label{sec:discussion}

Having obtained an explicit analytical form of the second-order
self-energy $\Sigma^{R}(\omega)$ at arbitrary frequency and bias
voltage, we now examine its various limiting behaviors. 
Most of those limiting behaviors have been known for the two-terminal
PH symmetric SIAM, so it's assuring to reproduce those expressions in
such a case.  Simultaneously, our results below provide
multi-terminal, PH asymmetric extensions of those asymptotic results.

\subsection{Equilibrium dot with and without the PH symmetry}

We can reproduce the equilibrium result by setting all the chemical
potentials equal, $\mu_{a\sigma} =
E_{d\sigma}=\epsilon_{d}+Un_{\bar{\sigma}}$. Then we immediately see
$\Xi_{3} = 0$ and
\begin{align}
& \Xi_{1}
 = 8 \cdot \big[ \tfrac{\pi^{2}}{4} (\tfrac{\varepsilon_{\sigma}}{i\gamma})
+ 3 \dilog(\tfrac{- \varepsilon_{\sigma}}{i\gamma}) \big],
\\
& \Xi_{2}
= 8\cdot \big[\frac{3\pi^{2}}{4} -
3\Lambda(\tfrac{\varepsilon_{\sigma} + 2i\gamma}{i\gamma})  
\big],
\end{align}
where $\varepsilon_{\sigma} = \omega - E_{d\sigma}$. As a result, 
the correlation part of the self-energy in equilibrium becomes
\begin{align}
 & \Sigma^{R}_{\sigma}(\omega) 
= \frac{i\gamma U^{2}}{\pi^{2} (\varepsilon_{\sigma} + i\gamma)}
  \bigg[ \frac{\tfrac{\pi^{2}}{4}(\tfrac{\varepsilon_{\sigma}}{i\gamma})
 + 3 \dilog(\tfrac{-\varepsilon_{\sigma}}{i\gamma})}{\varepsilon_{\sigma} - i\gamma} 
\nonumber \\ & \qquad \qquad \qquad \qquad
+  \frac{\tfrac{3\pi^{2}}{4} 
- 3 \Lambda (2 +
\tfrac{\varepsilon_{\sigma}}{i\gamma})}{\varepsilon_{\sigma}
+3i\gamma} 
\bigg].
\label{eq:Sigma-Equilibrium}
\end{align}
The PH symmetric case in particular corresponds to
$\varepsilon_{\sigma} = \omega$.  It reproduces the perturbation
results by Yamada and Yoshida~\cite{Yamada75,Yamada75b,Yoshida75} up
to the second-order of $U$, when we expand the above for small $\omega$.  The
PH symmetric result is indeed identical with the one obtained in
Ref.~\onlinecite{Muhlbacher11} for arbitrary frequency (See also
Eq.~\eqref{eq:PH-symmetric-SigmaR} below).

\subsection{Nonequilibrium PH symmetric dot connected with two terminals}

\citet{Muhlbacher11} have evaluated analytically 
the self-energy and the spectral function for the two-terminal PH symmetric
SIAM. In our notation, it
corresponds to the case $\gamma_{L}=\gamma_{R}=\gamma/2$, and
$E_{d\sigma}=0$.  When we parametrize the two chemical
potentials by $\mu_{a} = \zeta_{a\sigma}= a eV/2$ with $a=\pm 1$ in
Eqs.~\eqref{eq:Xi123}, the self-energy can be written as
\begin{align}
 & \Sigma^{R}_{\sigma}(\omega) = \frac{i\gamma U^{2}}{8\pi^{2}(\omega
 + i\gamma)}
  \left[ \frac{\Xi_{1}(\omega)}{\omega - i\gamma} +
    \frac{\Xi_{2}(\omega)}{\omega +3i\gamma} \right],
\label{eq:PH-symmetric-SigmaR}
\end{align}
where 
\begin{align*}
& \Xi_{1} = \tfrac{2\pi^{2} \omega}{i\gamma} 
+ 6 \sum_{a,b} \Big[ \dilog(\tfrac{- \omega + a eV /2}{b eV/2 +  i\gamma}) 
+ \tfrac{1}{4} \Log^{2} (\tfrac{- a eV/2 + i\gamma}{b eV/2 + i\gamma})
\Big], \\
& \Xi_{2}
=  6\pi^{2} 
- 6\sum_{a,b} \Big[   
 \Lambda(\tfrac{\omega -a eV/2 + 2i\gamma}{b eV/2 +  i\gamma}) 
+ \frac{1}{4} \Log^{2} (\tfrac{a eV/2 + i\gamma}{b eV/2 + i\gamma}) \Big].
\end{align*}
The above result are identical with what
Ref.~\onlinecite{Muhlbacher11} obtained.

\subsection{Expansion of small bias and frequency}

We now employ the small-parameter expansion of $\Sigma^{R}$ around the
half-filled equilibrium system.  Here we assume parameters
$\zeta_{a\sigma} = \mu_{a} - E_{d\sigma}$ and $\varepsilon_{\sigma} =
\omega - E_{d\alpha}$ are much smaller than the total relaxation rate
$\gamma$.
The expansion of $\Xi_{1}$ is found to contain the first- and
second-order terms regarding $\zeta_{a\sigma}$ and
$\varepsilon_{\sigma}$, while $\Xi_{2,3}$ do only the second-order terms.
Therefore the result of the expansion up to the second-order of these small
parameters is presented as
\begin{align}
\Sigma^{R}_{\sigma}(\omega) \approx
 \frac{iU^{2}}{8\pi^{2}\gamma}  \left[
 \Xi_{1} - \frac{\Xi_{2}}{3}
- \frac{\Xi_{3}}{2} \right].
\end{align}
Functions $\Xi_{i}$ can be expanded straightforwardly by using the
Taylor expansion of dilogarithm in Appendix C.  They are found to behave
\begin{align}
& \Xi_{1}(\varepsilon) \approx  8\, \Big[ 
\frac{(\pi^{2} -12) \varepsilon + 4\bar{\mu}}{4i\gamma} 
+ \frac{3 \varepsilon^{2} + 9 \overline{\mu^{2}}
 - 6 \varepsilon
  \bar{\mu} -  2 \bar{\mu}^{2}}{4(i\gamma)^{2}}
\big], \\
& \Xi_{2}(\varepsilon) \approx  - 8 \, 
\Big[  \frac{3( -\varepsilon^{2} + 2
  \varepsilon\bar{\mu} + 2
  \bar{\mu}^{2} - 
  \overline{\mu^{2}} - 2
  \overline{\mu^{2}})}{4(i\gamma)^{2}}  
\Big ]\\
& \Xi_{3} \approx \frac{16 \bar{\mu}^{2}}{(i\gamma)^{2}},
\end{align}
where $\bar{\mu}$ is defined in Eq.~\eqref{def:mubar} and we have
introduced 
\begin{align}
& \overline{\mu^{2}} 
= \sum_{a} \frac{\gamma_{a}}{\gamma} \mu_{a}^{2}
= \bar{\mu}^{2} + (\delta \mu)^{2}.
\end{align}
Combining all of these, we reach
the small bias/frequency behavior of the self-energy
$\Sigma^{R}_{\sigma}$ as
\begin{align}
& \Sigma^{R}_{\sigma}(\omega) 
\approx
 \frac{U^{2}}{\pi^{2} \gamma^{2}} \left[ (\tfrac{\pi^{2}}{4}
    - 3) (\omega - E_{d\sigma}) + \bar{\mu}\right]
\nonumber \\ & \qquad \qquad
-  \frac{iU^{2}}{2\pi^{2}\gamma^{3}} \Big[ 
(\omega - \bar{\mu})^{2}
+ 3\, (\delta \mu)^{2} \Big].
\label{eq:small-parameter-expansion}
\end{align}

Small bias expansion of $\Im \Sigma^{R}$ for the two-terminal system
has been discussed and determined by the argument based on the Ward
identity.~\cite{Oguri02} The dependence appearing in
Eq.~\eqref{eq:small-parameter-expansion} fully conforms to it (except
for the presence of the bare interaction instead of the renormalized
one).  Indeed, correspondence is made clear by noting
the parameters $\bar{\mu}$ and $\overline{\delta \mu^{2}}$ for the
two-terminal case become
\begin{align}
  & \bar{\mu} = \frac{\gamma_{L} \mu_{L} + \gamma_{R}
    \mu_{R}}{\gamma}; \quad
 (\delta \mu)^{2} = \frac{\gamma_{L} \gamma_{R}}{\gamma^{2}}
  (eV)^{2}.
\end{align}
The presence of linear term in $\omega$ and $V$ for the two-terminal
PH asymmetric SIAM was also emphasized recently.~\cite{Aligia12}

\subsection{Large bias voltage behavior}

One expects naively that the limit of large bias voltage $eV\to
\infty$ corresponds to the high temperature limit $T \to \infty$ in
equilibrium; it was shown to be so for the two-terminal PH symmetric
SIAM.~\cite{Oguri02}
We now show that the same applies to the multi-terminal
SIAM where bias voltages of the leads are pairwisely
large, \textit{i.e.}, half of them are positively large, and the others are
negatively large.

In the large bias voltage limit, all the arguments of dilogarithm
functions appearing in Eqs.~\eqref{eq:Xi123} become
$\pm 1$, where the values of dilogarithm are known (see
Appendix C).  Accordingly, the pairwisely large bias limit of
$\Xi_{i}$ is found to be
\begin{align}
& \Xi_{1}(\varepsilon) \approx \frac{2\pi^{2}(\varepsilon
  -i\gamma)}{i\gamma} , \\ 
& \Xi_{2}(\varepsilon) \approx -4\pi^{2}, \\
& \Xi_{3} \approx 0.
\end{align}
Correspondingly, the retarded self-energy becomes
\begin{align}
  & \Sigma^{R}_{\sigma}(\omega) \approx \frac{U^{2}/4}{\omega -
    E_{d\sigma} + 3i\gamma}
\end{align}
It shows that the result of the multi-terminal SIAM is the
same with that of the two-terminal PH symmetric SIAM except for a
frequency shift.  Accordingly, the retarded Green function
$G^{R}(\omega)$ in this limit is given by 
\begin{align}
  G^{R}_{\sigma}(\omega) 
\approx \frac{1}{\omega - E_{d\sigma} +i\gamma -
  \frac{U^{2}/4}{\omega - E_{d\sigma} + 3i\gamma}}.
\end{align}
The form indicates that for sufficiently strong interaction $U \gtrsim 2\gamma$, the dot
spectral function has two peaks at $E_{d\sigma} \pm U/2 = \epsilon_{d}
+ U (n_{\bar{\sigma}}\pm 1/2)$ with broadening $2\gamma$, so 
the system is driven into the the Coulomb blockade regime.  On the other hand, for weak
interaction $U < 2\gamma$, it has only one peak with two different values
of broadening that reduce to $\gamma$ and $3\gamma$ in the $U \to 0$
limit.

What is the role of the current preservation condition
Eq.~\eqref{eq:self-consistency} in this limit?  It just
determines the dot occupation number explicitly.  In fact, the
condition becomes
\begin{align}
& n_{\sigma} 
= \frac{-1}{\pi} \sum_{a} \frac{\gamma_{a}}{\gamma} \Im
\int^{(\mu_{a}-E_{d\sigma})/\gamma}_{-\infty}
\frac{dx}{x +i -\frac{u^{2}}{x + 3i}}  
\end{align}
with $u = U/(2\gamma)$, and $n_{\sigma}$ is independent of the
interaction strength because bias voltage sets the largest scale.
One can evaluate the above integral exactly to have
\begin{align}
& \int \frac{dx}{x +i -\frac{u^{2}}{x + 3i}} 
=  \sum_{s=\pm 1}
  \frac{\sqrt{1-u^{2}}+s}{2\sqrt{1-u^{2}}}  \Log(x-\alpha_{s}),
\end{align}
where $\alpha_{\pm} = -2i \pm i \sqrt{1-u^{2}}$. 
As a result, expanding it up to the second order of $u$ leads to
\begin{align}
& n_{\sigma} 
\approx \sum_{a} \frac{\gamma_{a}}{\gamma} \theta(\mu_{a})
- \frac{1}{\pi} \sum_{a} \frac{\gamma_{a}}{\mu_{a}}.
\end{align}
The first term corresponds to the occupation number that one expects
naturally from the effective distribution $\bar{f}$; it corresponds,
for instance, $\gamma_{L}/(\gamma_{L}+\gamma_{R})$ for the
two-terminal dot with $\mu_{R}< 0< \mu_{L}$ with $|\mu_{R,L}|\to
\infty$.  The second term is the deviation from it, which is
proportional to the average of the inverse chemical potential weighted
by the leads.

\section{Nonequilibrium spectral function}

\label{sec:spectral-fn}

We now turn our attention to the behavior of the nonequilibrium dot
spectral function, using our analytical expression of the self-energy
Eqs.~(\ref{eq:SigmaR},\ref{eq:Xi123}).  Below we particularly focus
our attention on the two cases: the two-terminal PH asymmetric SIAM
where current preservation has been an issue, and the multi-terminal
PH symmetric SIAM where the role of multiple leads has been
raising questions.
In all of the calculations below, we have checked numerically the
validity of the spectral weight sum rule at each configuration of bias
voltages. 

\subsection{Self-consistent current-preserving calculation}

\label{sec:numerics}


As was emphasized in Sec.~\ref{sec:current-formula}, when a dot system
does not retain the PH symmetry, the stationary current is not automatically
conserved and one must impose the current preservation condition
Eq.~\eqref{eq:self-consistency} explicitly.
As the right-hand side of Eq.~\eqref{eq:self-consistency} also depends
on the dot occupation number $n_{\sigma}$, this requires us to
determine $n_{\sigma}$ self-consistently by using the retarded Green
function in a certain approximation --- the second-order perturbation
theory in the present case.

\begin{figure}
  \centering
  \includegraphics[width=0.95\linewidth]{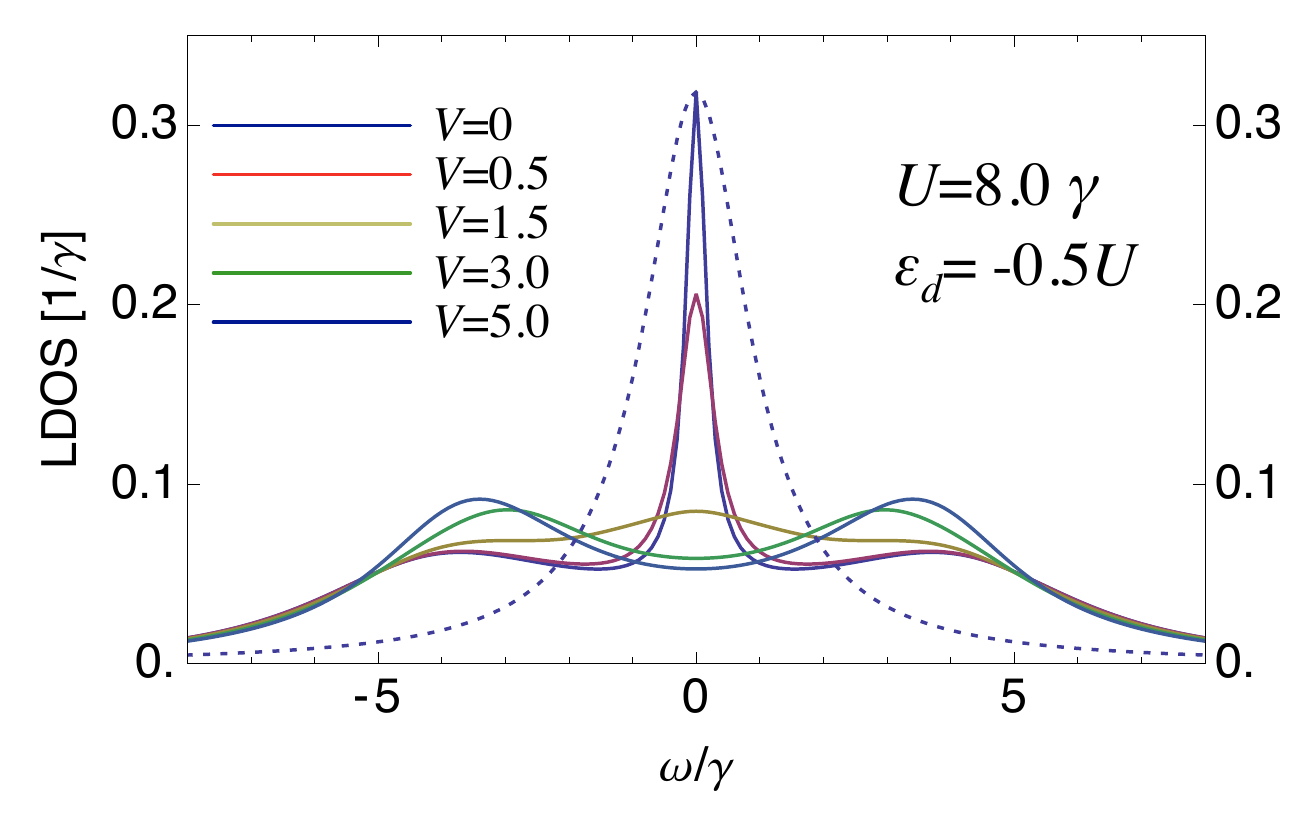}
  \caption{(Color online) Nonequilibrium dot spectral function of the
    two-terminal PH symmetric SIAM ($\epsilon_{d}=-U/2$) at finite
    bias voltage $eV = 0, 0.5, 
    1.5, 3.0, 5.0 \gamma$. The interaction strength is chosen as
    $U=8\gamma$. The dotted line represents the result of $U=0$ and $V=0$. 
  }
\label{fig:PH-symmetric}
\end{figure}

Figure~\ref{fig:PH-symmetric} shows the result of nonequilibrium dot
spectral function of the two-terminal PH symmetric SIAM at bias
voltage $eV = 0, 0.5, 1.5, 3.0$, and $5.0 \gamma$, which is
essentially the same result with Ref.~\onlinecite{Muhlbacher11} (of a
different set of parameters).  The occupation number is fixed to be
$n_{\sigma} = 1/2$ in this case, so its self-consistent determination
is unnecessary.  The results were compared favorably with those
obtained by diagrammatic quantum Monte Carlo
calculations~\cite{Muhlbacher11}; a relatively good quantitative
agreement was observed up to $U \sim 8\gamma$ (where the Bethe ansatz
Kondo temperature $k_{B} T_{K}= 0.055\gamma$~\cite{Tsvelick83b} while
the estimated half width of the Kondo resonance $k_{B}
\tilde{T}_{K}=0.23\gamma$) and bias voltage $V \lesssim 2\gamma$.
Applying bias voltage gradually suppresses the Kondo resonance without
splitting it, and the two peaks at $\pm U/2$ are developed at larger
bias voltages, which corresponds to the discussion in the previous
section.

\begin{figure}
  \centering
  \includegraphics[width=0.95\linewidth]{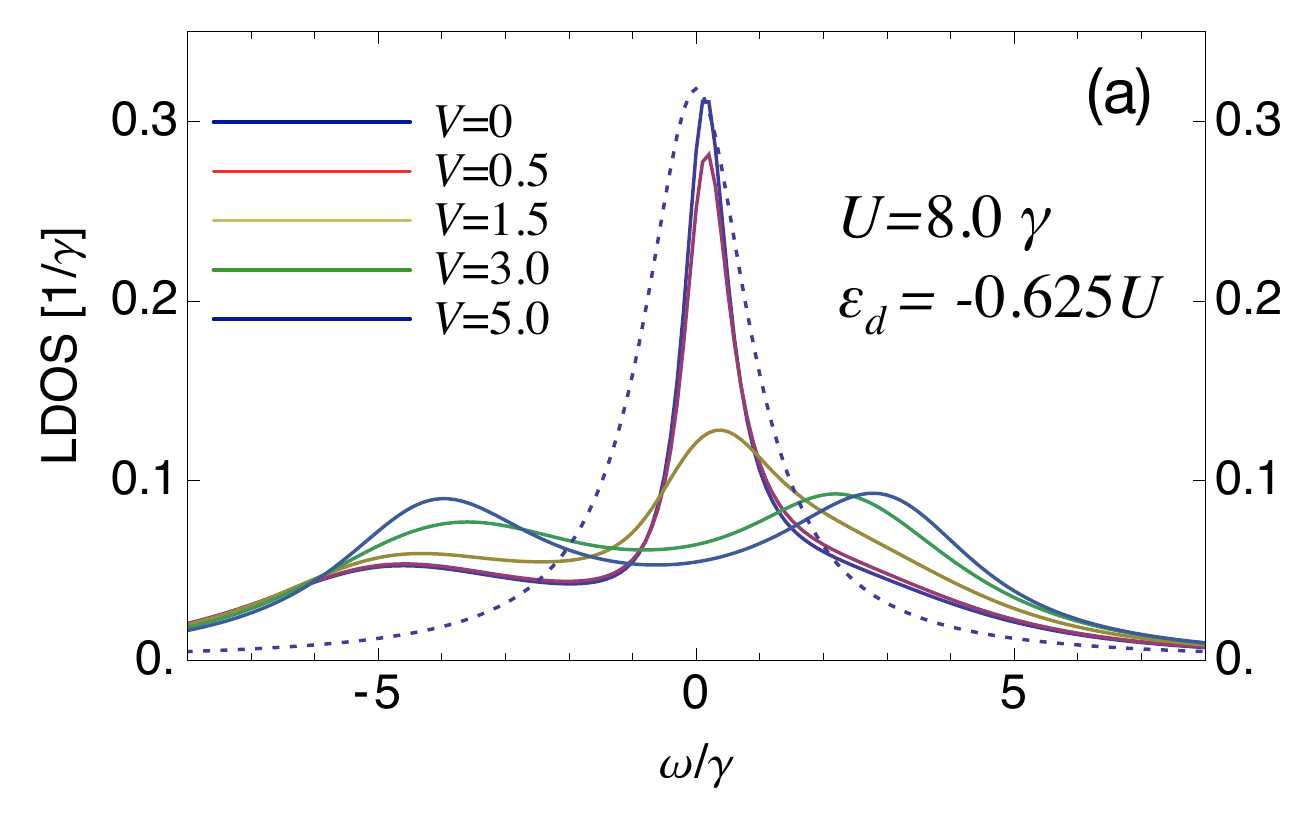}
  \includegraphics[width=0.95\linewidth]{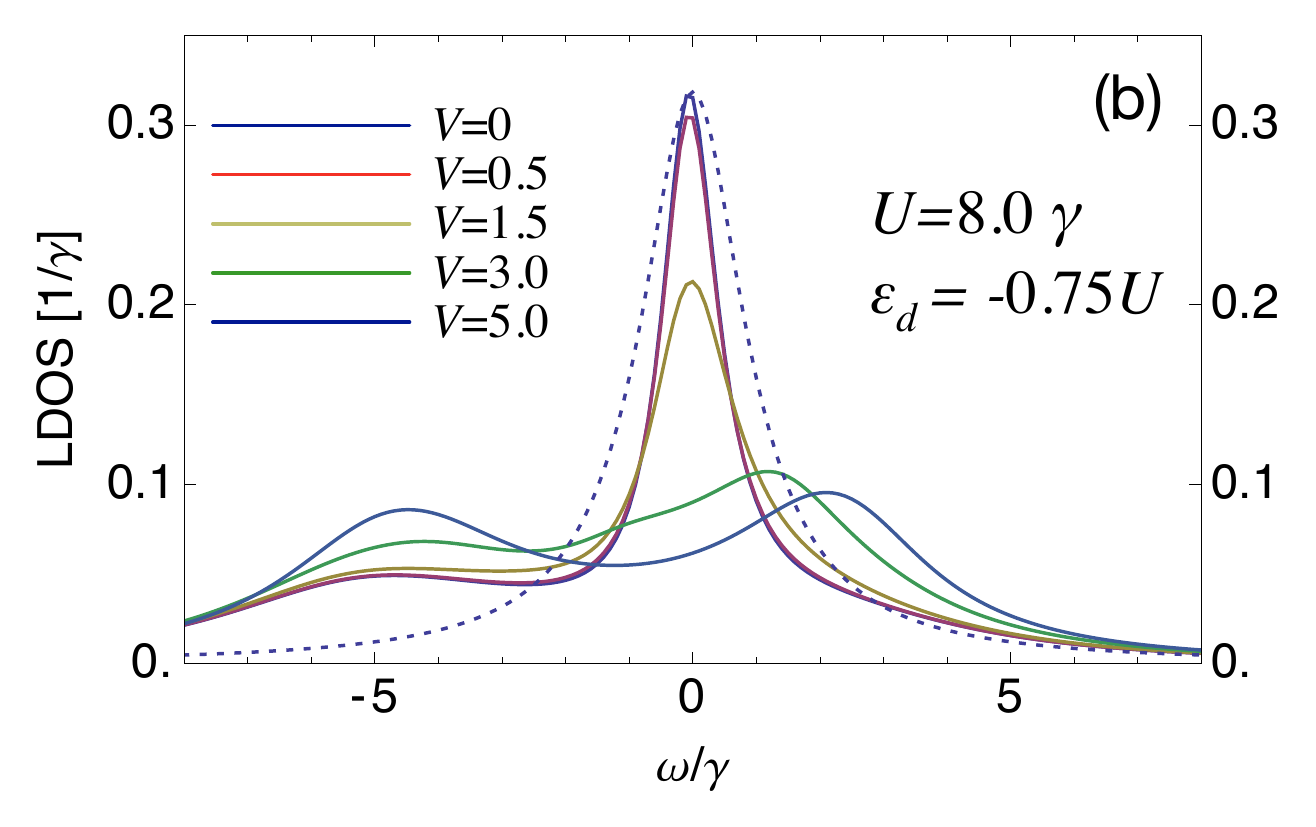}
  \caption{(Color online) Nonequilibrium dot spectral function of the
    PH asymmetric SIAM at (a) $\epsilon_{d} = -0.625U$, and (b)
    $\epsilon_{d} = -0.75U$.  All the other parameters are the same
    with Fig.~\ref{fig:PH-symmetric}. As an eye guide, the PH
    symmetric result of $U=0$ and $V=0$ is shown as a dotted line.}
\label{fig:PH-asymmetric}
\end{figure}

Figure~\ref{fig:PH-asymmetric} shows the result of our self-consistent
calculation of the nonequilibrium spectral function for the
two-terminal PH asymmetric SIAM at (a) $\epsilon_{d} = -0.625U$ and
(b) $\epsilon_{d}=-0.75U$.  A paramagnetic-type solution is assumed in
determining $n_{\sigma}$.  
As in the PH symmetric SIAM, one sees increasing bias voltage not
split but suppress the Kondo resonance while it develops a peak around
$E_{d} - U/2$.  The Kondo resonance peak is suppressed more
significantly at $\epsilon_{d} = -0.625U$ than at $-0.75U$, because
the Kondo temperature of the former ($k_{B}T_{K} \approx 0.067\gamma$;
$k_{B} \tilde{T}_{K} \approx 0.49\gamma$) is smaller than that of
the latter ($k_{B}T_{K} \approx 0.12\gamma$; $k_{B} \tilde{T}_{K}
\approx 0.58\gamma$).
An interesting feature of the PH asymmetric SIAM is that spectral
weight of the Kondo resonance seems shifting gradually toward $E_{d} +
U/2$ with increasing bias voltage, without exhibiting a three-peak
structure in the PH symmetric case.  This suggests a strong spectral
mixing between the Kondo resonance and a Coulomb peak at finite bias
voltage. Because of it, the interval of the two peaks at finite bias
is observed as roughly $U/2$ and gets widened up to $U$ for larger
$eV$.  
The bias dependence somehow looks similar to what was obtained by
assuming equilibrium noninteracting effective distribution for
$n_{\sigma}$~\cite{Matsumoto00} (which is hard to justify in our
opinion), though we emphasize our present calculation only relies on
the current-preservation condition without using any further assumption.
It is remarked that the effect shown by bias voltage is quite
reminiscent of finite temperature effect that was observed in the PH
asymmetric SIAM in equilibrium~\cite{Horvatic87}.  

\begin{figure}
  \centering
  \includegraphics[width=0.95\linewidth]{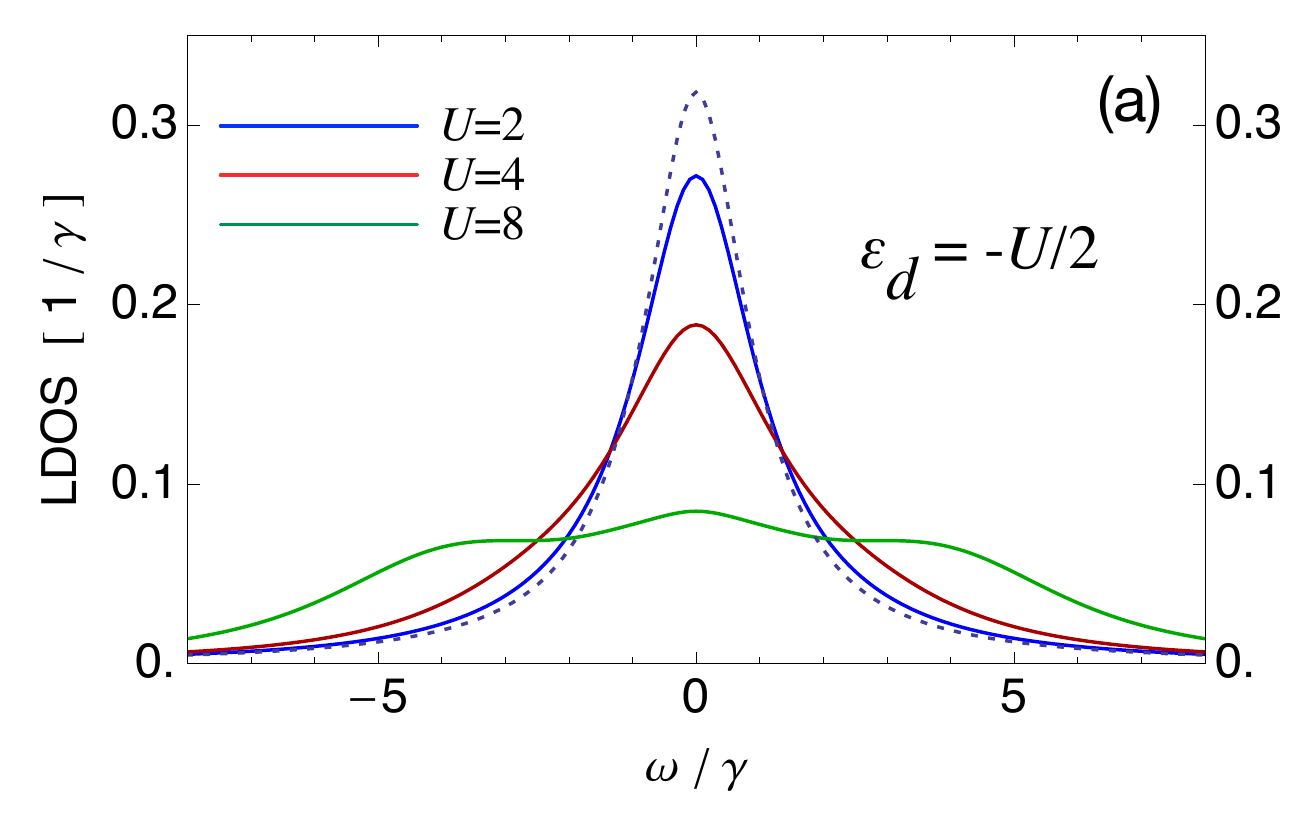}
  \includegraphics[width=0.95\linewidth]{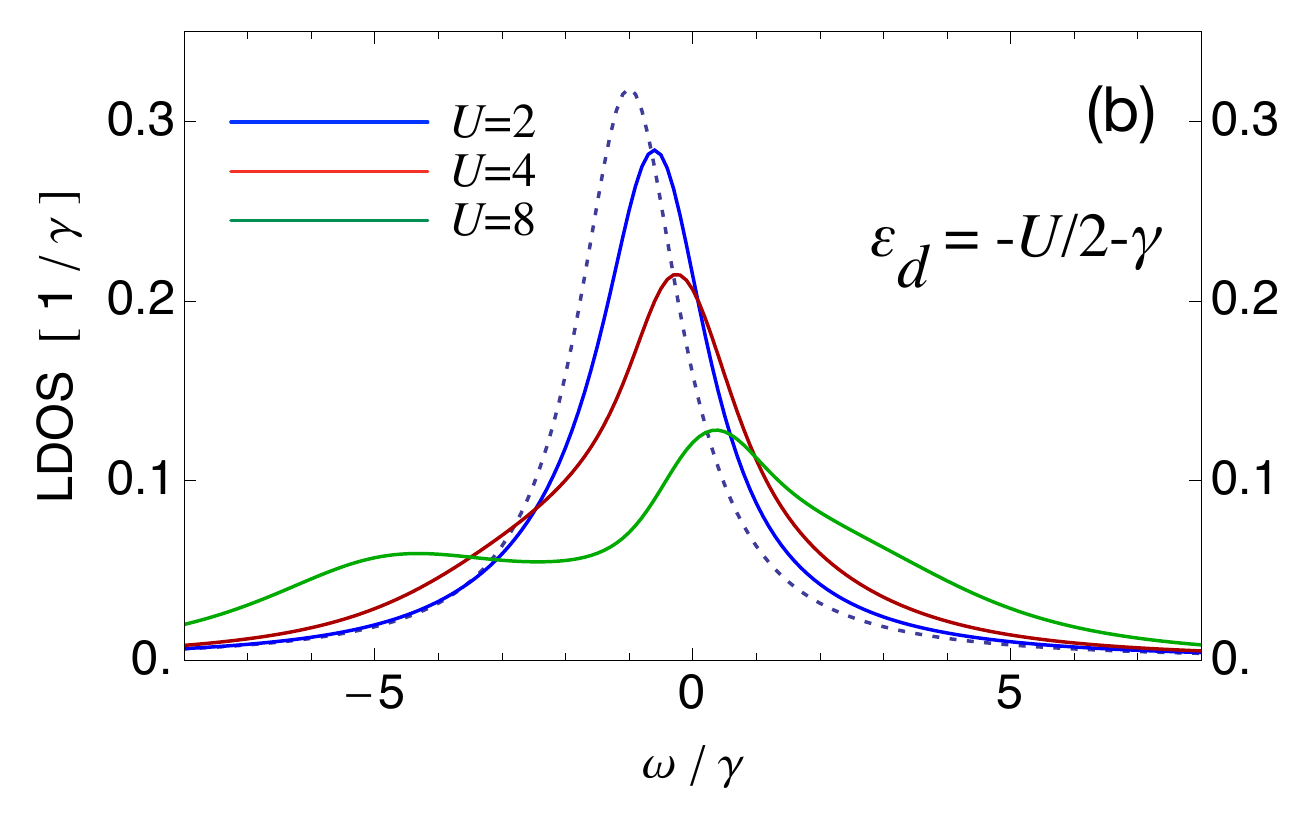}
  \caption{(Color online) Nonequilibrium dot spectral function for
    different values of interaction strength at bias voltage
    $eV=1.5\gamma$. Results of the interaction strengths $U = 2\gamma$, $4\gamma$,
    and $8\gamma$ are shown, while dotted lines refer to the
    noninteracting case as an eye guide. (a) spectral function of the PH symmetric
    SIAM at $\epsilon_{d} = -U/2$. (b) spectral function of the PH
    asymmetric SIAM at $\epsilon_{d} = -U/2 - \gamma$.}
\label{fig:LDOS-U}
\end{figure}

More insight can be gained by examining how the spectral structure
depends on the interaction strength at finite bias
voltage. Figures~\ref{fig:LDOS-U} (a,b) show a structural crossover
from a noninteracting resonant peak (the dotted line) to correlation
peaks, for (a) the PH symmetric SIAM $\epsilon_{d} = -U/2$, and (b)
the PH asymmetric SIAM $\epsilon_{d} = -U/2-\gamma$.  The PH symmetric
SIAM shows introducing $U$ leads to developing the correlation two
peaks as well as the Kondo peak that is suppressed by finite bias
voltage.  In contrast, the bias voltage effect on the PH
asymmetric SIAM is more involved, because the Kondo resonance is
apparently shifted and mixed with one of the correlation peaks,
eventually showing the two-peak structure at $E_{d} \pm U/2$ in large
bias voltage limit.

\subsection{Multi-terminal PH symmetric SIAM}

\begin{figure}
  \centering
  \includegraphics[width=0.95\linewidth]{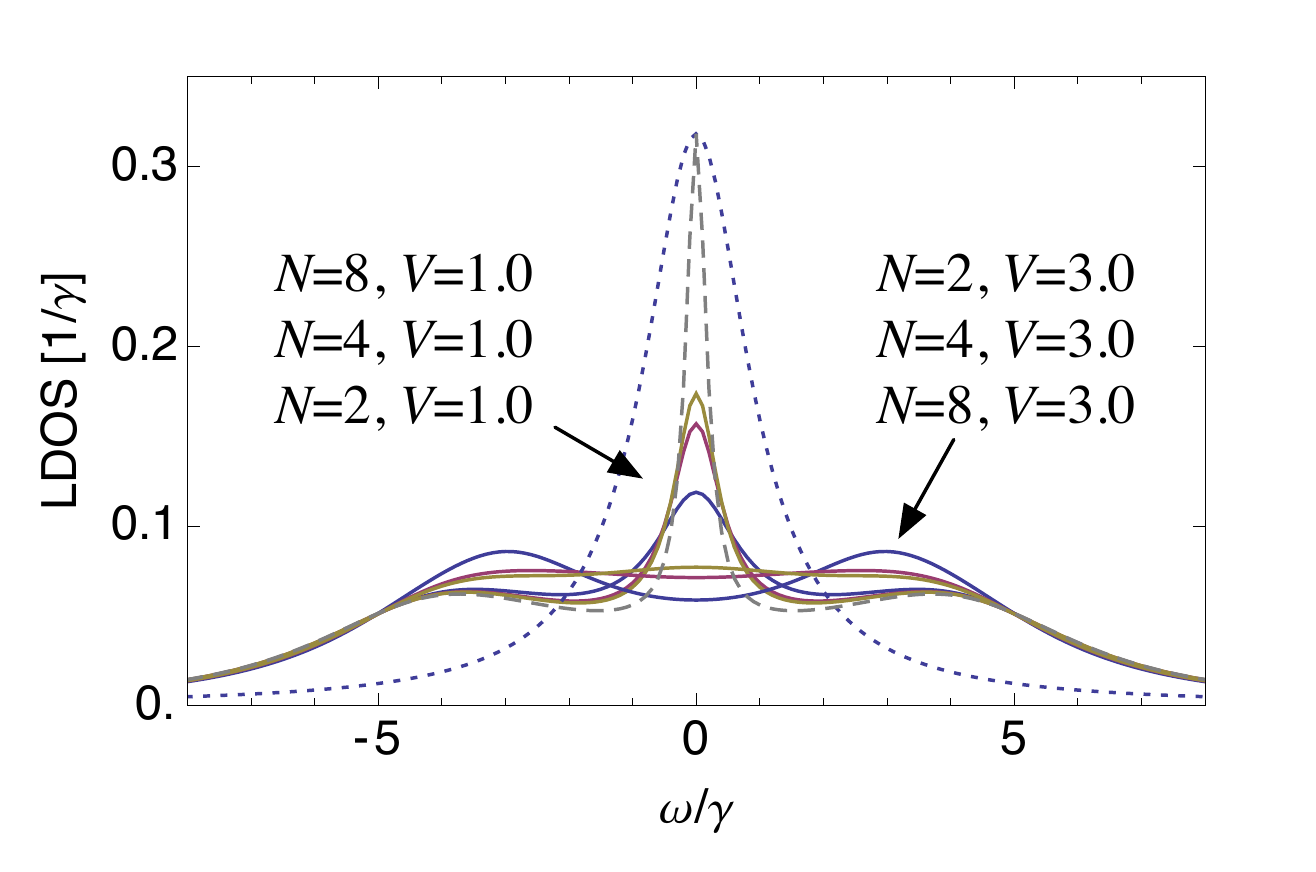}
  \caption{(Color online) Nonequilibrium dot spectral function for the
    PH symmetric multi-terminal dot ($N=2,4$, and $8$). Other
    parameters are chosen as the same as in
    Fig.~\ref{fig:PH-symmetric}. The dotted line correspond to the
    two-terminal noninteracting unbiased case, while the dashed line,
    to the two-terminal interacting unbiased case.}
\label{fig:NLDOS-by-V}
\end{figure}

To examine finite bias affects further and see particularly how the
presence of multi-terminals affects the nonequilibrium spectral
function, we configure a special setup of the multi-terminal SIAM that
preserves the PH symmetry: the dot is connected with $N$ identical
terminals, with bias levels being distributed
equidistantly between $-V/2$ and $+V/2$, and each of relaxation
rates is set to be $\gamma/N$.  The latter ensures that the unbiased
spectral function is the same, hence the Kondo temperature.
Results of the nonequilibrium spectral function are shown in
Fig.~\ref{fig:NLDOS-by-V}.  Again, we confirm that no splitting of the
Kondo resonance is observed in this multi-terminal setting.  One sees
further that the increasing the number of terminals \emph{enhances}
the Kondo resonance. This can be understood by weakening the bias
suppression effect on the Kondo resonance for a larger $N$. More
precisely, one may estimate the suppressing effect from small
bias behavior Eq.~\eqref{eq:small-parameter-expansion}.  Hence $\delta
\mu$ is a control parameter.  In the present multi-terminal PH
symmetric setting, the quantity $\delta \mu$ is found to be
\begin{align}
& \delta \mu = V \sqrt{\frac{N+1}{12(N-1)}}.
\end{align}
Therefore $\delta \mu$ decreases with increasing $N$, which results in
weakening the suppression and enhancing the Kondo resonance for a
larger $N$.

\begin{figure}
  \centering
  \includegraphics[width=0.95\linewidth]{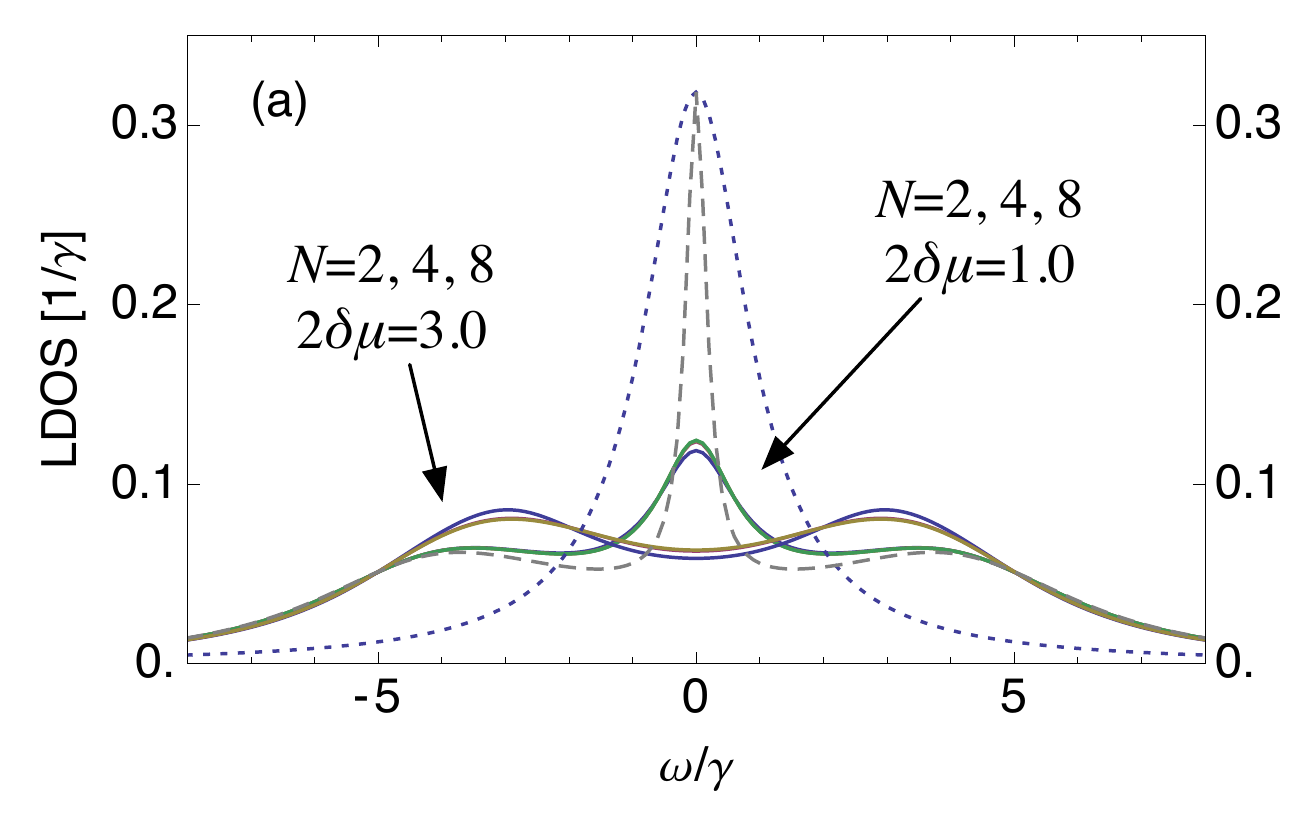}
  \includegraphics[width=0.95\linewidth]{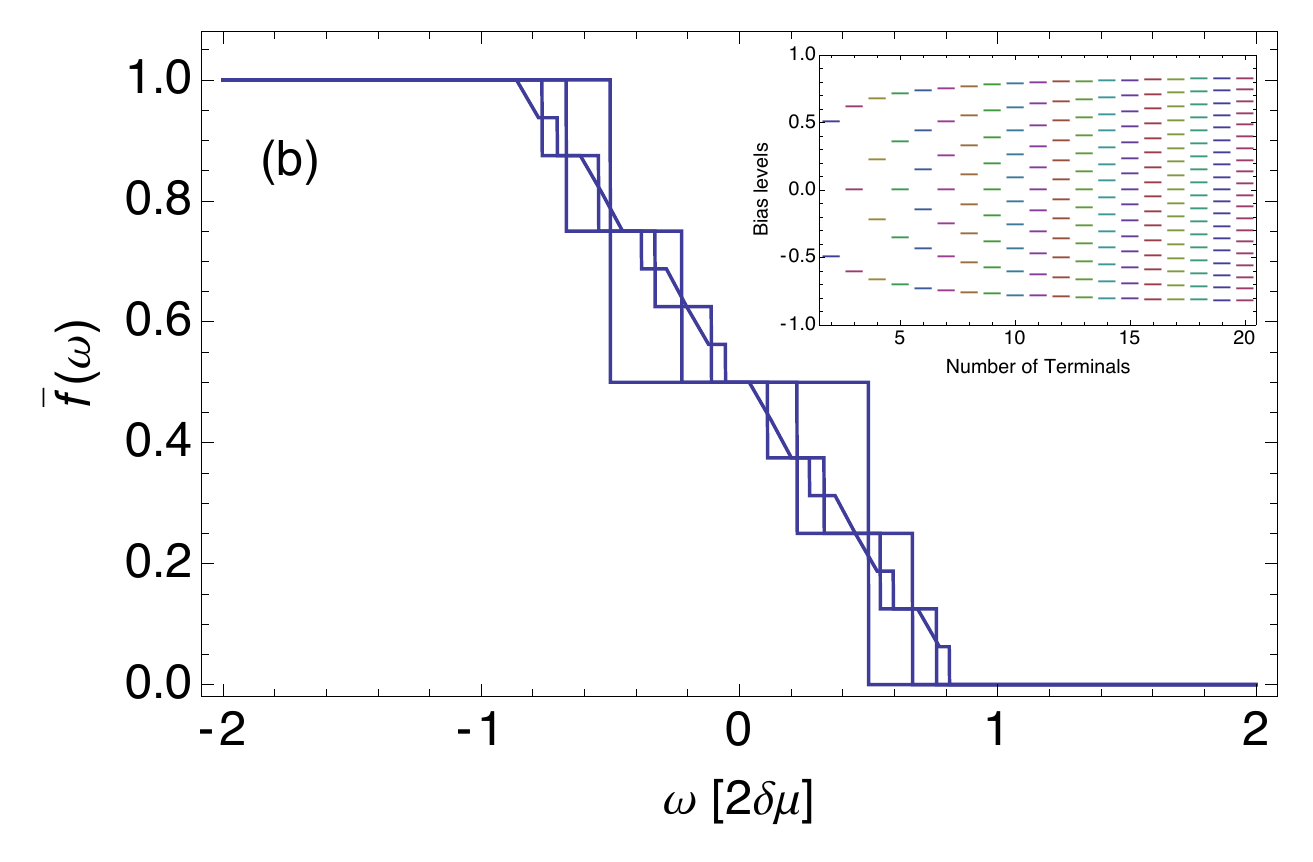}
  \caption{(Color online) (a)~Nonequilibrium spectral function for the
    PH symmetric multi-terminal dot with fixing $\delta \mu$
    ($N=2,4,8$). Other parameters are the same as in
    Fig.~\ref{fig:NLDOS-by-V}.  (b)~The effective Fermi distribution
    $\bar{f}(\omega)$ at zero temperature for the PH symmetric
    multi-terminal dot ($N=2,4,8,16$). The inset shows relative
    locations of bias levels with fixed $\delta\mu$ as a function of
    the number of terminals.}
\label{fig:NLDOS-by-dmu}
\end{figure}

The preceding argument also tells us that if the spectral function
bears any multi-terminal signatures at all, they would be more
conspicuous by examining it with fixing $\delta \mu$ rather than $V$.
This is done in Fig.~\ref{fig:NLDOS-by-dmu} (a);
Figure~\ref{fig:NLDOS-by-dmu} (b) shows how the effective dot
distribution $\bar{f}(\omega)$ and the relative locations of bias
levels (in the inset) evolve for a fixed $\delta \mu$ when $N$
increases. No multi-terminal signature in the nonequilibrium spectral
function is seen in Fig.~\ref{fig:NLDOS-by-dmu} (a); results of
different $N$ actually collapse, not only around zero frequency but in
the entire frequency range.  It suggests that the suppression of the
Kondo resonance deeply correlates the development of Coulomb peaks,
and a mixing between those spectral weight is important.  The
parameter $\delta \mu$ controls a crossover from the Kondo resonance
to the Coulomb blockade structure. We may also understand the
similarity between bias effect and temperature effect by the
connection through the large $N$ limit of the effective Fermi
distribution $\bar{f}(\omega)$, as shown in
Fig.~\ref{fig:NLDOS-by-dmu} (b).

\subsection{Finite bias effect on the spectral function: issues and speculation}

Although there is a consensus that bias voltage starts
suppressing the Kondo peak, and eventually destroys it with developing
the two Coulomb peaks when bias voltage is much larger than the Kondo
temperature, there is a controversy as to whether the Kondo resonance
peak will be split or not in the intermediate range of bias voltage.
All the results obtained by the second-order perturbation consistently
indicate that there is no split of the Kondo resonance; finite bias
voltage starts to suppress the Kondo resonance, and develops the two
Coulomb blockade peaks by shifting the spectral weight from the Kondo
resonance. 
We should mention that some other approximations draw a different
conclusion. Here we make a few remarks on apparent
discrepancy seen in various theoretical approaches as well as
experiments, as well as some speculation based on our results. 

Typically, several approaches that rely on the infinite $U$ limit,
notably non-crossing approximation, equation of motion method, and
other approaches investigating the Kondo Hamiltonian, observed the
splitting of the Kondo resonance under finite bias
voltage~\cite{Meir93,Wingreen94,Shah06}.  Those results, however, have
to be interpreted with great care, in our view.  Generally speaking,
the spectral function obtained by those approaches does not obey the
spectral weight sum rule: ignoring the doubly occupied state typically
leads to the sum rule $- \int^{\infty}_{-\infty} \Im
G^{R}_{\sigma}(\omega)/\pi = 1/2$~\cite{Meir93}, rather than the
correct value.  Therefore only half of the spectral weight can
be accounted for in those methods.  
Simultaneously, such (false) sum rule in conjugation with the
bias-suppression of the Kondo resonance cannot hep but lead to a
two-peak structure of the spectrum within the range of attention.
Splitting of the Kondo resonance might be an artifact of the
approximation.  Not fulfilling the correct sum rule, those approaches
may not be able to distinguish whether finite bias will split the
Kondo resonance or simply suppresses it with developing the Coulomb peaks.
As for the two-terminal PH symmetric SIAM, fourth-order contribution
regarding the Coulomb interaction $U$ has been evaluated
numerically~\cite{Fujii03,Fujii05,Hamasaki07}. The results seem
unsettled, though.  While \citet{Fujii03,Fujii05} suggested the fourth-order
term may yield the splitting of the Kondo resonance in the
intermediate bias region $k_{B}\tilde{T}_{K} \lesssim eV \lesssim U$
for sufficiently large interaction $U/\gamma \gtrsim 4$, which the
second-order calculation fails to report, another numerical study indicates
that the spectral function remains qualitatively the same with the
second-order result~\cite{Hamasaki07}.  
Experimental situation is not so transparent, either.  While the
splitting of the Kondo resonance was reported in a three-terminal
conductance measurement in a quantum ring system~\cite{Leturcq05}, a
similar spitting observed in the differential conductivity was
attributed to being caused by a spontaneous formation of ferromagnetic
contacts, not purely to bias effect~\cite{Nygard04}. It is also
pointed out that it has been recently recognized that the Rashba
spin-orbit coupling induces spin polarization nonmagnetically in a
quantum ring system with a dot when applying finite bias
voltage~\cite{Sun06,Crisan09,Taniguchi12}; hence such spin
magnetization might possibly lead to the splitting of the Kondo
resonance. 

The Kondo resonance is a manifestation of singlet formation between
the dot and the lead electrons.  One may naively think that when
several chemical potentials are connected with the dot, such singlet
formation would take place at each lead \emph{separately}, causing
multiple Kondo resonances.  The results of the multi-terminal PH
symmetric SIAM presented in the previous section tempt us to speculate
a different picture. Let us suppose that (almost) the same dot
distribution function $f_{\text{dot}}(\omega) =
G^{-+}(\omega)/(2i\pi)$ is realized for a fixed $\delta \mu$ with
a different terminal number $N$, as Fig.~\ref{fig:NLDOS-by-dmu} (a) suggests.  Note the
assumption is fully consistent to the Ward identity for low bias, but
it invalidates a quasiparticle ansatz $-\bar{f}(\omega) \Im
G^{R}(\omega)/\pi$ that explicitly depends on $N$.  
In the large $N$ limit with a fixed $\delta \mu$, the effective Fermi
distribution $\bar{f}(\omega)$ resembles the Fermi distribution at
finite temperature $k_{B}T \sim \delta\mu$.  Accordingly bias voltage
may well give effect similar to finite temperature. It is seen in low
and large bias limit for a dot with or without the PH symmetry. It
implies that a dot electron cannot separately form a singlet with the
lead at each chemical potential, because it needs to implicates states
at different chemical potentials through coupling with other
leads. Our second-order perturbation results seem to support this
view.

\section{Conclusion}
\label{sec:conclusion}

In summary, we have evaluated analytically the second-order
self-energy and Green function for a generic multi-terminal single
impurity Anderson model in nonequilibrium.  Various limiting behaviors
have been examined analytically.  Nonequilibrium spectral function
that preserves the current is constructed and is checked to satisfy
the spectral weight sum rule.  The multi-terminal effect is examined
for the PH-symmetric SIAM, particularly.  Within the validity of the
present approach, it is shown that the Kondo peak is not split due to
bias voltage.  It is found that most of the finite bias effect is
similar to that of finite temperature in low- and high-bias limits
with and without the PH symmetry. Such nature could be understood by
help of the Ward identity and the connection through the $N \gg 1$
terminal limit.
The present analysis serves as a viable tool that can cover a wide
range of experimental situations. Although there is still a chance
that high-order contributions might generate a new effect such as
split Kondo resonances in a limited range of parameters, it is
believed that the second-order perturbation theory can capture the
essence of the Kondo physics in most of realistic situations.
Moreover, having a concrete analytical form that satisfies both the
current conservation and the sum rule, the present work provides a
good, solid, workable result that more sophisticated future treatment
can base on.


\begin{acknowledgments}
  The author gratefully acknowledges A.~Sunou for fruitful
  collaboration that delivered some preliminary results in this
  work.  The author also appreciates R.~Sakano and A.~Oguri for
  helpful discussion at the early stage of the work. The work was
  partially supported by Grant-in-Aid for Scientific Research (C)
  No.~22540324 and 26400382 from MEXT, Japan.
\end{acknowledgments}

\appendix

\label{sec:appendix}

\section{Noninteracting Green functions with finite bias}

We start with the nonequilibrium Green function $G$ without the
Coulomb interaction on the dot.  Its Keldysh structure is specified by 
\begin{align}
G_{\sigma}(\omega)
= \begin{pmatrix}
  \omega - \epsilon_{d} + i \gamma (1-2\bar{f}) & 
  + 2i \gamma \bar{f} \\
  -2 i \gamma (1-\bar{f}) & 
 \!\!\!\!\!\!\! - (\omega - \epsilon_{d}) + i\gamma (1-2\bar{f})
\end{pmatrix}^{-1},
\end{align}
where $\bar{f}$ is the effective dot distribution defined in
Eq.~\eqref{eq:effective-f}.  We incorporate the Hartree-type diagram
into the unperturbed part by replacing
$\epsilon_{d}$ to $\epsilon_{d} \mapsto E_{d\sigma} = \epsilon_{d} + U
n_{\bar{\sigma}}$. Note $n_{\bar{\sigma}}$ is the exact dot
occupation that needs to be determined consistently later. 
After mapping to the RAK representation, those components are given by
\begin{align}
& G^{R,A}_{\sigma}(\omega) = \frac{1}{\omega - E_{d\sigma}
    \pm i\gamma}, \\
&  G^{K}_{\sigma} (\omega) = [1-2\bar{f}(\omega)]
  \left[ G^{R}_{\sigma} (\omega) - G^{A}_{\sigma} (\omega) \right].
\label{eq:GK0}
\end{align}
The function 
$1-2\bar{f}(\omega)$ reduces to $\sum_{a} (\gamma_{a}/\gamma) \sgn (\omega -
\mu_{a})$ at zero temperature.

\section{Nonequilibrium polarization part}

Taking the Fourier transformation of Eqs.~\eqref{eq:def-Pi}, using
Eq.~\eqref{eq:GK0}, and making further manipulations, we can rewrite
$\Pi^{R}$ and $\Pi^{K}$ as
\begin{align}
\Pi^{R}(\varepsilon)
&= \sum_{a} \frac{\gamma_{a}}{\gamma} \frac{\gamma B_{aa}
  (\varepsilon)}{\pi \varepsilon (\varepsilon+2i\gamma)} = \left[
  \Pi^{A}(\varepsilon) \right]^{*}, 
\label{eq:PiR}\\ 
\Pi^{K}(\varepsilon) &= 
2i\sum_{a,b} \frac{\gamma_{a} \gamma_{b}}{\gamma^{2}}  \coth
    \tfrac{\beta(\varepsilon-\mu_{ab})}{2} 
 \Im \left[ \frac{\gamma
    B_{ab}(\varepsilon)}{\pi \varepsilon (\varepsilon+2i\gamma)}
\right],
\label{eq:PiK}
\end{align}
where $\mu_{ab} = \mu_{a} - \mu_{b}$, $\beta$ is the inverse
temperature, and $B_{ab}(\varepsilon)$ is given by 
\begin{align}
& B_{ab}(\varepsilon) 
=  \int\!\! d\varepsilon' \!
\left[ f_{b}(\varepsilon') - f_{a}(\varepsilon'+\varepsilon)  \right]
\!\left[
  G^{A}(\varepsilon') - G^{R}(\varepsilon'+\varepsilon) \right].
\end{align}
In the present work, we are interested in the zero temperature limit,
for which $\coth (\beta x)$ becomes $\sgn(x)$.  The function $B_{ab}$
in this limit is evaluated as (with $\zeta_{a\sigma} = \mu_{a} - E_{d\sigma}$)
\begin{align}
&B_{ab}(\varepsilon) 
= -\log \left( \frac{\varepsilon - \zeta_{a\sigma} +
    i\gamma} {-\zeta_{b\sigma} + i\gamma} \right) 
- \log\left( \frac{\varepsilon + \zeta_{b\sigma} +
    i\gamma}{\zeta_{a\sigma} + i\gamma} \right).
\end{align}
This corresponds to a multi-terminal extension of the result obtained
for the two-terminal PH symmetric SIAM.

\section{Dilogarithm with a complex variable}

To complete evaluating the remaining integral over $E$ of 
Eqs.~(\ref{eq:I1},\ref{eq:I2}), we take full advantage of various
properties of dilogarithm function $\dilog(z)$. A concrete integral
formula we have utilized will be given in Appendix D.  For the
sake of completeness, we here collect its definition and properties
necessary to complete our evaluation.

\paragraph*{Definition}

Dilogarithm $\dilog(z)$ with a complex argument $z \in \mathbb{C}$ is
not so commonly found in literature.  As it is a multi-valued
function, we need to specify its branch structure properly.
One way to define dilogarithm $\dilog(z)$ all over the complex plane
consistently is to use the integral representation
\begin{align}
& \dilog (z) = - \int^{z}_{0}dt \, \frac{\Log(1-t)}{t}.
    \label{def:dilog}
\end{align}
The multivaluedness of dilogarithm $\dilog$ originates from logarithm
in the integrand.  Here we designate the principal value of logarithm as
$\Log$, which is defined by
\begin{align}
    \Log z = \ln |z| + i \Arg z; \quad \text{(for $-\pi < \Arg z \le
      \pi$)}.
    \label{def:Log-of-z}
\end{align}
According to Eq.~\eqref{def:dilog}, $\dilog(z)$ has a branch-cut just
above the real axis of $x>1$. Accordingly, its values just above and
below the real axis are different for $x>1$: $\dilog(x-i\eta)
=\dilog(x)$ but $\dilog(x+i\eta) =\dilog(x) + 2i\pi \ln x$.  Some
special values are known analytically. We need $\dilog(0) = 0$,
$\dilog(1) = \pi^{2}/6$, $\dilog(-1) = -\pi^{2}/12$ and $\dilog(2) =
\pi^{2}/4 - i \pi \ln 2$ for evaluation later.

\paragraph*{Functional relations}
\label{sec:functional-relations}

Dilogarithm $\dilog(z)$ has interesting symmetric properties regarding
its argument $z$; values at $z$, $1-z$, $1/z$, $1/(1-z)$, $(z-1)/z$
and $z/(z-1)$ are all connected with one another.  
Those points are ones generated by symmetric operations $S$
and $T$ defined by
\begin{align}
& S z = \frac{1}{z}; \quad T z = 1-z,
\end{align}
and $\{I, S, T, ST, TS, TST \}$ forms a group. Other operations correspond
to 
\begin{align}
& ST z = \frac{1}{1-z}; \quad TS z = \frac{z-1}{z},\\
&TST z  = STS z = \frac{z}{z-1}.
\end{align}
Applying a series of integral transformations in
Eq.~\eqref{def:dilog}, one can connect the values of dilogarithm at
these values with one another~\cite{Maximon03}.  Note, those
functional relations are usually presented only for real arguments.
Extending them for complex variables needs examining its branch-cut
structure carefully.  By following and extending the derivations in
Ref~\onlinecite{Maximon03} for complex $z \in \mathbb{C}$, we prove
that the following functional relations are valid for any complex
variable $z$.
\begin{align}
&\dilog(Sz) = - \dilog(z) -\tfrac{\pi^{2}}{6} -
    \tfrac{1}{2} \left[ \Log(TSz) - \Log(Tz) \right]^{2}, 
\label{eq:dilog-func-rel-start}\\
&\dilog(Tz) =
    -\dilog (z) + \tfrac{\pi^{2}}{6} - \Log(Tz) \Log z, \\
&\dilog(TSTz) = - \dilog(z) -\tfrac{1}{2}
    \Log^{2}(STz) 
\nonumber \\ & \qquad \qquad \qquad
- [ \Log(Tz) + \Log(STz) ] \Log z, \\
&\dilog\big( TSz \big) = \dilog(z) -
    \tfrac{\pi^{2}}{6} 
\nonumber \\ & \qquad \qquad \qquad
-\tfrac{1}{2} \Log^{2}(Sz) 
- \Log(Sz) \Log(Tz), \\
&\dilog \big(STz \big) = \dilog(z) +
    \tfrac{\pi^{2}}{6} 
\nonumber \\ & \qquad \qquad \qquad
+ \tfrac{1}{2} \Log^{2}(Tz) + \Log(Tz) \Log(TSTz).
\label{eq:dilog-func-rel-end}
\end{align}
To our knowledge, the above form of extension of
functional relations of dilogarithm has not been found in literature.

\paragraph*{The Taylor expansion}

To examine various limiting behaviors, we need the Taylor expansion of
dilogarithm, which is derived straightforwardly from
Eq.~\eqref{def:dilog},
\begin{align}
  & \dilog(z) = \dilog(z_{0}) - \sum_{k=1}^{\infty}
  \frac{(z-z_{0})^{k}}{k!}  \frac{d^{k-1}}{dz^{k-1}} 
    \frac{\Log(1-z)}{z} \bigg|_{z=z_{0}}.
\end{align}
The presence of $\Log(1-z)$ reflects the branch-cut structure of
$\dilog(z)$.  In particular, we utilize the following expansion in
our analysis.
\begin{align}
& \dilog(z) \approx z +
  \frac{z^{2}}{4} + \frac{z^{3}}{9} + \frac{z^{4}}{16} + \cdots, \\
&    \Lambda(2+z) \approx \frac{\pi^{2}}{4} -\frac{z^{2}}{4} +
\frac{z^{3}}{6} - \frac{5 z^{4}}{48} + \cdots. 
\end{align}

\section{Integral formula}

Here we derive and present the central integral formula for 
evaluating Eqs.~(\ref{eq:I1},\ref{eq:I2}).
By performing a simple integral transformation in
Eq.~\eqref{def:dilog}, we have the integration
\begin{align}
&\int^{z}_{-b} \frac{\Log \left(\tfrac{x+b}{c}\right)}{x-a} dx 
= \int^{\frac{z+b}{a+b}}_{0} \frac{\Log \left( \tfrac{a+b}{c} y
  \right)}{y-1} dy
\\ &  \qquad
= \Log
\left( \tfrac{z+b}{c} \right)\, \Log \left( 1- 
  \tfrac{z+b}{a+b} \right) + \dilog\left(
  \tfrac{z+b}{a+b} \right),
\end{align}
where all the parameters $(a,b,c)$ as well as $z$ may be taken as complex
numbers.
Combined with fractional decomposition, we see the
following integral be evaluated in terms of dilogarithm:
\begin{align}
& \int^{z}_{-b} \frac{\Log \left( \tfrac{x+b}{c} \right) \,
    dx}{(x-a_{1})(x-a_{2})(x-a_{3})}
\nonumber \\ & \quad
= \sum_{i=1}^{3} \frac{\Log \left( \tfrac{z+b}{c} \right) \Log \big(
  1-\frac{z+b}{a_{i}+b} 
    \big) + \dilog \left( \tfrac{z+b}{a_{i}+b}\right)}{\prod_{j\neq i}
    (a_{i}-a_{j})}.
\label{eq:integral-formula}
\end{align}

\section{Calculation of the correlated part of the self-energy $\Sigma_{\sigma}$}

The remaining task to complete calculating $\Sigma^{R}$ in the form of
Eqs.~(\ref{eq:SigmaR}--\ref{eq:Xi123}) is to
collect all the relevant formulas and organize them in a form that
conforms to Eq.~\eqref{eq:integral-formula}.  
To write concisely, we introduce the following notations
\begin{align}
& \mu_{ab} = \mu_{a} - \mu_{b}, \\
& \zeta_{a\sigma} = \mu_{a} - E_{d\sigma}, \\
& \varepsilon_{\sigma} = \omega - E_{d\sigma}. 
\end{align}
where the Hartree level $E_{d\sigma}$ is defined as before.
We express the terms $I_{1}$ and $I_{2}$ defined in
Eqs.~(\ref{eq:I1},\ref{eq:I2}) as
\begin{align}
 & I_{1}  = -\sum_{a,b}\sum_{\alpha,\beta=\pm 1} \frac{\alpha\gamma_{a}
    \gamma_{b}}{\pi\gamma} 
\nonumber \\ & \qquad \times
\int^{+\infty}_{-\infty}dE\, \, \frac{\,\sgn
    (E-\beta\mu_{ab})}{(E+\varepsilon_{\sigma}   + i\gamma) }
  \frac{\Log \left( \tfrac{E -\beta \zeta_{a\bar{\sigma}} + i\alpha
        \gamma}{-\beta \zeta_{b\bar{\sigma}} +  i\alpha \gamma}
    \right)}{(E+2i\alpha \gamma)E}, \\
& I_{2} = -\sum_{a,b}\sum_{\alpha,\beta=\pm 1}
\frac{\alpha \gamma_{a}\gamma_{b}}{\pi\gamma}
\nonumber \\ & \qquad \times
\int^{+\infty}_{-\infty} dE 
\frac{\sgn(E+\varepsilon_{\sigma}-\zeta_{a\sigma})}{(E+\varepsilon_{\sigma} +
  i\alpha \gamma) } \frac{\Log \left(\frac{E + 
    \beta \zeta_{b\bar{\sigma}} - i\gamma}{\beta \zeta_{b\bar{\sigma}} - i\gamma}
\right)}{E(E-2i\gamma)}. 
\end{align}
Here the leads $a,b$ in $I_{1}$ as well as $b$ in $I_{2}$ carry spin
$\bar{\sigma}$, while $a$ in $I_{2}$ does spin $\sigma$.  Singularity
on energy integration is prescribed by the principal values.  Equation
\eqref{eq:integral-formula} enables us to perform and express the above
integrals in terms of dilogarithm.  The resulting expressions are
still complicated, but we can simplify them further using functional
relations of dilogarithm
Eqs~(\ref{eq:dilog-func-rel-start}--\ref{eq:dilog-func-rel-end}).  These
require straightforward but rather laborious manipulations.  In this
way, we reach the final expression of $\Sigma^{R}_{\sigma}$ of
Eq.~\eqref{eq:SigmaR}.


\end{document}